\documentclass[showpacs,nofootinbib,showkeys,eqsecnum,prd,aps,twocolumn,twoside]{revtex4-1}
\usepackage{amsfonts,amsmath,amssymb,bm,graphicx}

\begin{document}

\title{Creation of Dirac neutrinos in a dense medium with a time-dependent
effective potential}
\author{Maxim Dvornikov${}^{a,b,c}$}
\email{maxim.dvornikov@anu.edu.au}
\author{S.~P.~Gavrilov${}^{a,d}$}
\email{gavrilovsergeyp@yahoo.com}
\author{D.~M.~Gitman${}^{a,e}$}
\email{gitman@if.usp.br}
\date{\today }

\affiliation{${}^{a}$Institute of Physics, University of S\~{a}o Paulo, CP 66318, CEP
05315-970 S\~{a}o Paulo, SP, Brazil;\\
${}^{b}$Research School of Physics and Engineering, Australian National
University, Canberra ACT 2601, Australia;\\
${}^{c}$Pushkov Institute of Terrestrial Magnetism, Ionosphere and Radiowave
Propagation (IZMIRAN), \\
142190 Troitsk, Moscow, Russia;\\
${}^{d}$Department of General and Experimental Physics, Herzen State
Pedagogical University of Russia, Moyka embankment 48, 191186
St.~Petersburg, Russia\\
$^{e}$Department of Physics, Tomsk State University, 634050, Tomsk, Russia}

\begin{abstract}
We consider Dirac neutrinos interacting with background fermions in the
frame of the standard model. We demonstrate that a time-dependent effective
potential is quite possible in a protoneutron star (PNS) at certain stages
of its evolution. For the first time, we formulate a nonperturbative
treatment of neutrino processes in a matter with arbitrary time-dependent
effective potential. Using linearly growing effective potential, we study
the typical case of a slowly varying matter interaction potential. We
calculate differential mean numbers of $\nu \bar{\nu}$ pairs created from
the vacuum by this potential and find that they crucially depend on the
magnitude of masses of the lightest neutrino eigenstate. These distributions
uniformly span up to $\sim 10~$eV energies for muon and tau neutrinos
created in PNS core due to the compression just before the hydrodynamic
bounce and up to $\sim 0.1~\mathrm{eV}$ energies for all three active
neutrino flavors created in the neutronization. Considering different stages
of the PNS evolution, we derive constraints on neutrino masses, $m_{\nu
}\lesssim (10^{-8}-10^{-7})\,\text{eV, }$corresponding to the nonvanishing $%
\nu \bar{\nu}$ pairs flux produced by this mechanism. We show that one can
distinguish such coherent flux from chaotic fluxes of any other origin. Part
of these neutrinos, depending on the flavor and helicity, are bounded in the
PNS, while antineutrinos of any flavor escape the PNS. If the created pairs
are $\nu _{e}\bar{\nu}_{e}$, then a part of the corresponding neutrinos also
escape the PNS. The detection of $\nu $ and $\bar{\nu}$ with such low
energies is beyond current experimental techniques.
\end{abstract}

\maketitle

\section{Introduction\label{S1}}

Particle creation from the vacuum by strong electromagnetic, Yang Mills, and
gravitational fields is a well-known nonlinear quantum phenomenon which has
many applications in modern high-energy physics. Its theoretical study has a
long story that is described in numerous works, see for example Refs.~\cite%
{GMR85,BirDav82,Gitman,GavGT06}. Creation of charged particles from the
vacuum by strong electric-like fields needs superstrong field magnitudes
compared with Schwinger critical field $E_{\mathrm{cr}}=m^{2}c^{3}/e\hbar
\simeq 1.3\times 10^{16}\,\mathrm{V}\cdot \mathrm{cm}^{-1}$ \cite{schwinger}%
. Nevertheless, recent progress in laser physics allows one to hope that
this effect will be experimentally observed in the near future even in
laboratory conditions, see Ref.~\cite{Dun09} for the review.\footnote{%
Electron-hole pair creation from the vacuum was recently observed in
graphene, see, for example, Ref.~\cite{GavGitY12}.} The particle creation
from the vacuum by external electric and gravitational backgrounds plays
also an important role in cosmology and astrophysics~\cite{BirDav82}.

It should be noted that not only electric and gravitational macroscopic
backgrounds may destabilize a quantum field vacuum. As it was shown in Ref.~%
\cite{GavGit13a}, the vacuum of neutrinos, possessing anomalous magnetic
moments, becomes unstable in a strong inhomogeneous magnetic field such that
the creation of neutrinos by the latter field may take place. Estimates
presented in Ref.~\cite{GavGit13a} show that this effect can be produced by
strong magnetic fields of magnetars and fields generated during a supernova
explosion and has to be taken into account in the astrophysics.

The instability of the neutrino vacuum exists also due to the neutrino
interaction with a background matter. It should be noted that the
neutrino-antineutrino ($\nu \bar{\nu}$) pairs creation in a dense matter of
a neutron star was studied in Refs.~\cite%
{Loeb90,Kach98,KiersW97,KusPos02,Koe05}. In Refs.~\cite%
{Loeb90,Kach98,KiersW97} the matter density was supposed to be
time-independent and the $\nu \bar{\nu}$ pair creation was considered
empirically by using the analogy between a neutron star potential and a
potential well. In this case the production rate of the $\nu \bar{\nu}$ pair
creation was evaluated semiclassically borrowing the Schwinger's result in
QED for the probability for a vacuum to remain a vacuum~\cite{schwinger}.
The case of a time-dependent density was studied nonperturbatively, using
numerical calculations, for an oscillating density of a neutron star, a
supernova, and gamma ray bursts in Ref.~\cite{KusPos02} and perturbatively
in Ref.~\cite{Koe05}. It should be noted that the perturbation theory is
valid only for nonrealistic high frequency density variations. Realistic $%
\nu \bar{\nu}$\ pairs creation due to a slowly varying matter interaction
potential was not considered before.

In the present article we formulate a consistent nonperturbative approach
for calculating, in the framework of QFT, the\emph{\ }$\nu \bar{\nu}$\emph{\ 
}pair production from the vacuum due to a coherent neutrino interaction with
a background matter, in particular, a matter with arbitrary time-dependent
effective potential. We apply then this approach to calculate the effect in
some interesting cases of the medium evolution and distribution.

The article is organized as follows. In the beginning we describe a field
theory model, which is used by us to treat neutrinos interacting with a
background matter. Then, in the framework of the quantum version of the
model, we consider a case of a matter with time-dependent effective
potential. We show that such a background is quite possible at certain
stages of a protoneutron star (PNS) evolution. For instance, one can discuss
the matter compression in the PNS core just before the hydrodynamic bounce
or the phase transition of a dense medium of PNS at the neutronization
stage. Then, using a nonperturbative approach that is similar to the one
developed in QED with time-dependent external electromagnetic fields, see
Ref.~\cite{Gitman}, we formulate a calculation scheme for the neutrino
production in the case under consideration. This technique is based on using
complete sets of exact solutions of a modified Dirac equation for neutrinos
interacting with a matter density. These solutions are used to quantize the
neutrino field and introduce the corresponding $\mathrm{in}$- and \textrm{out%
}- creation and annihilation operators. We represent the mean numbers of $%
\nu \bar{\nu}$ pairs created and probabilities of all the transitions via
coefficients of the corresponding Bogolyubov transformations. In particular,
we derive general formulas that describe the $\nu \bar{\nu}$ pair creation
in the matter with linearly growing in time effective potential and study
the typical case of a slowly varying matter interaction potential.

As a main application of the developed approach, we consider the $\nu \bar{%
\nu}$ pair creation of Dirac neutrinos from the vacuum due to the
compression in the core of PNS before the bounce and at the neutronization
stage. We show that the behavior of the effective number density at these
stages of the PNS evolution can be described by a slowly varying in time
homogeneous effective potential. Then we demonstrate that the intensity of
the neutrino creation crucially depends on\textbf{\ }the magnitude of masses
of the lightest neutrino eigenstate. We also find that the momentum
distribution of $\nu \bar{\nu}$ pairs is isotropic and uniform in the
low-energy range (up to $\sim 10~\mathrm{eV}$)\ dropping sharply for higher
energies. We find that if the mass of the lightest neutrino is small enough,
the flux of pairs of the lightest $\nu $ and $\bar{\nu}$, created from the
vacuum during the stages of PNS evolution, may exceed the low-energy flux of
any other origin. We derive constraints on neutrino masses corresponding to
the nonvanishing $\nu \bar{\nu}$ pairs flux produced from the vacuum due to
the compression in the PNS before the bounce and at the neutronization
stage. Finally, we list all the obtained results. Possible accompanying
processes that might affect identification of this vacuum instability at the
initial stages of the PNS evolution are examined in Appendix~\ref{SS5.2}.
Some mathematical details are separated in Appendix~\ref{App}.

\section{Interaction of Dirac neutrinos with background matter\label{S2}}

Here we briefly consider the classical field theory description of massive
Dirac neutrinos interacting with background fermionic matter.

The results of the recent experiments (see, e.g., Ref.~\cite{An12})
explicitly demonstrate that neutrinos are massive particles and there is a
nonzero mixing between different mass eigenstates. However, in some cases
one can neglect the mixing in the neutrino sector. For example, it is the
case when the corresponding transition probability of neutrino oscillations
is suppressed. In such cases we can consider a single neutrino eigenstate
having an effective mass $m$. It should be noted that the question whether
neutrinos are Dirac or Majorana particles still remains open (see, e.g.,
Ref.~\cite{Ago13}). In our constructions and further calculations we work
with Dirac neutrinos. We suppose that the gravitational interaction of
neutrinos is negligible and the effect of possible matter rotation is small
for quantum processes under consideration.

The Lagrangian of a massive Dirac neutrino field $\psi \left( X\right) $
interacting with a matter by an effective potential $g_{\mu }\left( X\right) 
$ has the following form in the~forward scattering approximation\footnote{%
Here we use the natural units in which $\hslash =c=1$.}%
\begin{align}
\mathcal{L}= & \bar{\psi}\left( X\right) \left( \mathrm{i}\gamma ^{\mu
}\partial _{\mu }-m\right) \psi \left( X\right)
\notag
\\
& -g_{\mu }\left( X\right) 
\bar{\psi}\left( X\right) \gamma ^{\mu }P_{\mathrm{L}}\psi \left( X\right) ,
\label{eq:LagrDir}
\end{align}%
see Ref.~\cite{Lagr97}. Here $\psi \left( X\right) $ is a Dirac spinor, $%
X=\left( x^{0}=t,\mathbf{r}=(x,y,z)\right) $,$\ \gamma ^{\mu }=\left( \gamma
^{0},\boldsymbol{\gamma }\right) $ are Dirac matrices, $\gamma ^{5}=\mathrm{i%
}\gamma ^{0}\gamma ^{1}\gamma ^{2}\gamma ^{3}$, and $P_{\mathrm{L}%
}=(1-\gamma ^{5})/2$ is the projector to the left chiral states . In what
follows, we use the Dirac matrices in the standard representation, 
\begin{gather}
\gamma ^{0}=\left( 
\begin{array}{cc}
1 & 0 \\ 
0 & -1%
\end{array}%
\right) ,\quad \boldsymbol{\gamma }=\left( 
\begin{array}{cc}
0 & \boldsymbol{\sigma } \\ 
-\boldsymbol{\sigma } & 0%
\end{array}%
\right) ,
\notag
\\
\gamma ^{5}=\left( 
\begin{array}{cc}
0 & 1 \\ 
1 & 0%
\end{array}%
\right) ,  \label{eq:Dirmat}
\end{gather}%
where $\boldsymbol{\sigma }$ are the Pauli matrices.

The effective potential $g^{\mu }\left( X\right) $ that describes the matter
interaction with neutrinos is a linear combination of the hydrodynamic
currents $j_{f}^{\mu }$ and polarizations $\lambda _{f}^{\mu }$ of
background fermions $f$, 
\begin{equation}
g^{\mu }\left( X\right) = \sqrt{2}G_{\mathrm{F}}\sum_{f}\left(
q_{f}^{(1)}j_{f}^{\mu }+q_{f}^{(2)}\lambda _{f}^{\mu }\right) ,
\label{efPot}
\end{equation}%
where $G_{\mathrm{F}}$ is the Fermi constant and coefficients $q_{f}^{(1)}$
and $q_{f}^{(2)}$ depend on the types of a neutrino and background fermions 
\cite{DvoStu02}. If we deal with electron neutrinos $\nu _{e}$ propagating
in the matter that is composed of electrons, protons, and neutrons, these
coefficients have the form, 
\begin{align}
q_{f}^{(1)}= & I_{\mathrm{L}3}^{(f)}-2Q_{f}\sin ^{2}\theta _{\mathrm{W}}+\delta
_{ef},
\notag
\\
q_{f}^{(2)}=&-I_{\mathrm{L}3}^{(f)}-\delta _{ef},
\label{eq:q1q2nue}
\end{align}%
where $I_{\mathrm{L}3}^{(f)}$ is the third component of the weak isospin of
the type $f$ fermions, $Q_{f}$ is their electric charge, $\theta _{\mathrm{W}%
}$ is the Weinberg angle, and $\delta _{ef}=1$ for electrons and vanishes
for protons and neutrons. To get the coefficients $q_{f}^{(1,2)}$ for muon
and tau neutrinos $\nu _{\mu ,\tau }$ we should set $\delta _{ef}$ to be
zero in Eq.~(\ref{eq:q1q2nue}).

Let us consider first an electroneutral matter which is unpolarized and
nonmoving. In this case the only zeroth component $g\left( X\right) \equiv
g^{0}\left( X\right) $ of $g^{\mu }\left( X\right) $ is nonzero. Using Eq.~(%
\ref{eq:q1q2nue}), this component can be found in the following form, 
\begin{align}
g\left( X\right) =&\sqrt{2}G_{\mathrm{F}}n_{\mathrm{eff}},
\notag
\\
n_{\mathrm{eff%
}}=&%
\begin{cases}
n_{e}-\frac{1}{2}n_{n}, & \text{for}\ \nu _{e}, \\ 
-\frac{1}{2}n_{n}, & \text{for}\ \nu _{\mu ,\tau },%
\end{cases}
\label{eq:gexpl}
\end{align}%
where $n_{e}$ and $n_{n}$ are the electron and neutron densities
respectively. The difference in the effective potentials for $\nu _{e}$ and $%
\nu _{\mu ,\tau }$ in Eq.~(\ref{eq:gexpl}) is owing to the fact that,
besides neutral current interactions, $\nu _{e}$ is also involved in the
charged current interactions with the given matter.

The Lagrangian (\ref{eq:LagrDir}) implies the following equations of motion, 
\begin{equation}
\left( \mathrm{i}\gamma ^{\mu }\partial _{\mu }-m-g\left( X\right) \gamma
^{0}P_{\mathrm{L}}\right) \psi \left( X\right) =0.  \label{eq:Direq}
\end{equation}%
In general case the effective potential depends on all the space-time
coordinates $X$. In the following we shall restrict ourselves to the case
when $g\left( X\right) $ is homogeneous and depends only on the time $t$.

This model can be applied for the description of neutrinos in realistic
conditions like a dense matter of PNS. Note that the matter of PNS with the
high degree of accuracy can be taken as spatially homogeneous~\cite%
{KeiJanMul96}. At certain stages of the supernova explosion the effective
potential can be regarded as a function of time only. For example, just
before the hydrodynamic bounce the matter density in PNS core increases
several orders of magnitude. Another situation when the effective potential
can be time dependent happens outside the core at the neutronization stage.
Indeed, a typical PNS has $n_{n}\approx n_{e}\approx n_{p}$ before the
neutronization. We can take that $n_{e}=n_{p}\approx 0$ in some regions
outside the PNS core after the neutronization. Therefore, using Eq.~%
\eqref{eq:gexpl}, we get that the value $g$ varies from the initial $g(t_{%
\mathrm{in}})$ to the final $g(t_{\mathrm{out}})$ as%
\begin{equation}
g(t_{\mathrm{out}})=%
\begin{cases}
-2g(t_{\mathrm{in}}) & \text{for}\ \nu _{e} \\ 
+2g(t_{\mathrm{in}}) & \text{for}\ \nu _{\mu ,\tau }%
\end{cases}%
.  \label{1.4}
\end{equation}%
Thus the time-dependent effective potential is quite possible in PNS. As is
demonstrated below, it is the time dependence of $g$ which stipulates the
instability of the neutrino vacuum and results in a coherent $\nu \bar{\nu}$
pairs creation.

One can see that the inhomogeneity of PNS matter near the star surface
affects the neutrino motion in the PNS crust and somehow influences the
neutrino creation. This effect requires a separate consideration. We shall
briefly discuss it in Appendix~\ref{SS5.2}.

Since $g$ is uniform, we can choose the Dirac spinor in the following form: 
\begin{equation}
\psi \left( X\right) =\exp \left[ -\frac{\mathrm{i}}{2}\int_{t_{0}}^{t}g(t^{%
\prime })dt^{\prime }\right] \tilde{\psi}\left( X\right) ,  \label{1.5}
\end{equation}%
where the spinor $\tilde{\psi}\left( X\right) $ satisfies the equation 
\begin{align}
\mathrm{i}\partial _{0}\tilde{\psi}\left( X\right) =&H\left( t\right) \tilde{%
\psi}\left( X\right) ,
\notag
\\
H\left( t\right) =&\gamma ^{0}\left( -\mathrm{i}%
\nabla \boldsymbol{\gamma }+m\right) -\frac{1}{2}g\left( t\right) \gamma
^{5}.  \label{eq:weDirHamf}
\end{align}%
One can see that the time-dependent Hamiltonian $H\left( t\right) $ is the
kinetic energy operator. Note that the Dirac Hamiltonian that corresponds to
the untransformed Eq.~(\ref{eq:Direq}) is $H\left( t\right) +g\left(
t\right) /2$. However, the Hamiltonian $H\left( t\right) $ plays an
important role in the physical interpretation of states vectors. It should
be also noted that in our case when $\nabla g=0$, both the momentum operator 
$-\mathrm{i}\nabla $ and the helicity operator, 
\begin{equation}
\Xi =\frac{-\mathrm{i}\nabla \boldsymbol{\Sigma }}{\sqrt{\left( -\mathrm{i}%
\nabla \right) ^{2}}},\quad \boldsymbol{\Sigma }=\gamma ^{5}\gamma ^{0}%
\boldsymbol{\gamma }=\left( 
\begin{array}{cc}
\boldsymbol{\sigma } & 0 \\ 
0 & \boldsymbol{\sigma }%
\end{array}%
\right) ,  \label{1.5.1}
\end{equation}%
commute with $H\left( t\right) $.

Using Eq.~(\ref{1.5}) we can verify that the inner product of arbitrary
solutions $\psi $ and $\psi ^{\prime }$ is reduced to the inner product of
the corresponding solutions $\tilde{\psi}$ and $\tilde{\psi}^{\prime },$ 
\begin{equation}
\left( \psi ,\psi ^{\prime }\right) =\int \psi ^{\dag }\left( t,\mathbf{r}%
\right) \psi ^{\prime }\left( t,\mathbf{r}\right) d\mathbf{r=}\left( \tilde{%
\psi},\tilde{\psi}^{\prime }\right) ,  \label{1.6}
\end{equation}%
and is conserved.

In what follows, we assume that $m\neq 0$. In this case $\gamma ^{5}$ does
not commute with the Hamiltonian $H\left( t\right) $. Then, using the
representation%
\begin{equation}
\tilde{\psi}\left( X\right) =\left[ \mathrm{i}\partial _{0}+H\left( t\right) %
\right] \phi \left( X\right) ,  \label{1.7}
\end{equation}%
we obtain the second-order differential equation for the spinor $\phi \left(
X\right) $, 
\begin{equation}
\left\{ \left( \partial _{0}\right) ^{2}+\left[ H\left( t\right) \right]
^{2}+\frac{\mathrm{i}}{2}\gamma ^{5}\partial _{0}g\left( t\right) \right\}
\phi \left( X\right) =0.  \label{1.8}
\end{equation}%
In particular, this equation describes the influence of the time dependence
of $g\left( t\right) $ on neutrino wave functions.

If $m=0$, the matrix $\gamma ^{5}$ commutes with $H\left( t\right) $ and
Eq.~(\ref{eq:Direq}) can be separated into two independent equations, 
\begin{align}
\mathrm{i}\partial _{0}\psi _{\mathrm{L,R}}\left( X\right) =&\left[
H_{0}\left( t\right) +\frac{1}{2}\left( 1\pm 1\right) g\left( t\right) %
\right] \psi _{\mathrm{L,R}}\left( X\right) ,
\notag
\\
H_{0}\left( t\right)
=&\gamma ^{0}\left( -\mathrm{i}\nabla \boldsymbol{\gamma }+m\right) ,
\label{eq:psiLR}
\end{align}%
for the spinors $\psi _{\mathrm{L,R}}\left( X\right) =\frac{1}{2}(1\mp
\gamma ^{5})\psi \left( X\right) $. Equation (\ref{eq:psiLR}) is a
first-order differential equation with respect of time. The spinors $\psi _{%
\mathrm{R}}\left( X\right) $ and%
\begin{equation}
\exp \left[ +\mathrm{i}\int_{t_{0}}^{t}g(t^{\prime })dt^{\prime }\right]
\psi _{\mathrm{L}}\left( X\right) ,
\end{equation}%
describe free neutrinos and antineutrinos since the potential $g\left(
t\right) $ is absent in equations for these quantities. Of course, it is a
consequence of our supposition that $g\left( X\right) $ is uniform. If,
however, $\nabla g\left( X\right) \neq 0$, the left neutrinos are not free
anymore. Hence the scale of the possible matter inhomogeneity $L$ has to be
big enough, e.g., $L\gg 1/m$.

\section{Quantization in terms of adequate particles and antiparticles\label%
{S3}}

In this section we use results of the canonical quantization of the
Lagrangian in Eq.~(\ref{eq:LagrDir}) described in Ref.~\cite{DvoGit13}. We
start with the constant and uniform effective potential. Then we consider
the matter with time-dependent effective potential. Using the corresponding
exact solutions of the Dirac equation, we introduce creation and
annihilation operators which diagonalize the kinetic energy operator. The
latter operator has a positive spectrum either in the initial or in the
final time instants. We construct the initial and final Fock spaces and
physical quantities that will be calculated in what follows.

\subsection{Constant effective potential\label{SS3.1}}

We start with the case when $g=\ $\textrm{const\ }$\neq 0$. Here the
one-particle description is possible, such that one can speak about one
neutrino moving in a homogeneous matter with a constant effective potential.
Then the Hamiltonian $H\left( t\right) =H$ is time independent. The
corresponding solutions of the Dirac equation are plane waves $\psi \left(
X\right) \sim \exp (-\mathrm{i}p_{\mu }X^{\mu }).$ Particles in such states
have the following kinetic energies $\mathcal{E}$~\cite{StuTer05}, 
\begin{equation}
\mathcal{E}=\sqrt{m^{2}+\left( p-\sigma \frac{g}{2}\right) ^{2}},
\label{eq:enlevDir}
\end{equation}%
where $p=|\mathbf{p}|$, $\mathbf{p}$ is the neutrino momentum, and $\sigma
=\pm 1$ is the eigenvalue of the neutrino helicity operator given by Eq.~(%
\ref{1.5.1}). The total energies $p_{0}^{\left( \pm \right) }$ differ from
the kinetic energies by a constant value, $p_{0}^{\left( \pm \right) }=\pm 
\mathcal{E+}g/2,$ since the density $g$ is homogeneous.

We represent wave functions under consideration as follows, 
\begin{align}
_{+}\psi (t,\mathbf{r})\sim & u_{\sigma }(\mathbf{p})\exp \left[ -\mathrm{i}%
p_{0}^{\left( +\right) }t+\mathrm{i}\mathbf{pr}\right] ,
\notag
\\
_{-}\psi (t,%
\mathbf{r})\sim & v_{\sigma }(\mathbf{p})\exp \left[ -\mathrm{i}p_{0}^{\left(
-\right) }t+\mathrm{i}\mathbf{pr}\right] ,  \label{psipmrqm}
\end{align}%
where the basis spinors $u_{\sigma }(\mathbf{p})$ and $v_{\sigma }(\mathbf{p}%
)$ have the form 
\begin{align}
u_{\sigma }=\sqrt{\frac{m+\mathcal{E}}{2\mathcal{E}}}\left( 
\begin{array}{c}
w_{\sigma } \\ 
\frac{\sigma p-g/2}{m+\mathcal{E}}w_{\sigma }%
\end{array}%
\right) ,
\notag
\\
v_{\sigma }=\sqrt{\frac{m+\mathcal{E}}{2\mathcal{E}}}\left( 
\begin{array}{c}
-\frac{\sigma p-g/2}{m+\mathcal{E}}w_{\sigma } \\ 
w_{\sigma }%
\end{array}%
\right) ,  \label{eq:basspinDir}
\end{align}%
and $w_{\sigma }=w_{\sigma }(\mathbf{p})$ are the two-component helicity
amplitudes (see Ref.~\cite{BerLifPit80}). These spinors satisfy the
following orthonormality conditions and completeness relations: 
\begin{eqnarray}
&&u_{\sigma }^{\dagger }(\mathbf{p})u_{\sigma ^{\prime }}(\mathbf{p})=\delta
_{\sigma \sigma ^{\prime }},\quad v_{\sigma }^{\dagger }(\mathbf{p}%
)v_{\sigma ^{\prime }}(\mathbf{p})=\delta _{\sigma \sigma ^{\prime }},
\notag
\\
&&u_{\sigma }^{\dagger }(\mathbf{p})v_{\sigma ^{\prime }}(\mathbf{p})=0, 
\notag \\
&&\sum_{\sigma }\left[ u_{\sigma }(\mathbf{p})\otimes u_{\sigma }^{\dagger }(%
\mathbf{p})+v_{\sigma }(\mathbf{p})\otimes v_{\sigma }^{\dagger }(\mathbf{p})%
\right] =1.  \label{eq:sumheli}
\end{eqnarray}

It is important to note that in the framework of the quantum field theory,
taking into account the fermion nature of neutrinos, one can see that $%
_{+}\psi (t,\mathbf{r})$ describes neutrino states with the kinetic energy $%
p_{0}^{\left( +\right) }-g/2=\mathcal{E},$ while $_{-}\psi (t,\mathbf{r})$
describes antineutrino states with the kinetic energy $\left\vert
p_{0}^{\left( -\right) }-g/2\right\vert =\mathcal{E}$. One can also see that
the corresponding neutrinos and antineutrinos behave like free particles.

\subsection{Time-dependent effective potential\label{SS3.2}}

In the case of a time-dependent effective potential $g\left( t\right) $, the
Hamiltonian $H\left( t\right) $ is also time dependent, and $H\left(
t\right) $ and $H\left( t^{\prime }\right) $ do not commute if $t\neq
t^{\prime }$. Using our experience in QED with external time-dependent
backgrounds, we believe that the one-particle description is not applicable
in such a case. To consider nonperturbative effects, we have to use the
approach developed in QED and known as the generalized Furry representation
(see Refs.~\cite{Gitman,GavGT06}). Below, we show that the problem in
question can be treated in the similar manner.

After the quantization, $\psi \left( X\right) =\psi \left( t,\mathbf{r}%
\right) $ turns out to be the Heisenberg operator $\Psi (X)=\Psi \left( t,%
\mathbf{r}\right) $. This operator obeys both the Dirac equation [Eq.~(\ref%
{eq:weDirHamf})] and the standard equal time anticommutation relations:%
\begin{align}
& \left[ \Psi \left( t,\mathbf{r}\right) ,\Psi \left( t,\mathbf{r}^{\prime
}\right) \right] _{+}=\left[ \Psi ^{\dag }\left( t,\mathbf{r}\right) ,\Psi
^{\dag }\left( t,\mathbf{r}^{\prime }\right) \right] _{+}=0,
\notag 
\\
& \left[ \Psi
\left( t,\mathbf{r}\right) ,\Psi ^{\dag }\left( t,\mathbf{r}^{\prime
}\right) \right] _{+}=\delta \left( \mathbf{r-r}^{\prime }\right) .
\label{3.2}
\end{align}%
The second quantized Hamiltonian $\hat{H}$ and the corresponding momentum
and helicity operators have the following forms: 
\begin{align}
\hat{H}\left( t\right) =&\int \Psi ^{\dagger }\left( t,\mathbf{r}\right)
H\left( t\right) \Psi \left( t,\mathbf{r}\right) d\mathbf{r+}H_{0}\left(
t\right) \mathbf{\ },  \label{3.3} \\
\mathbf{\hat{p}}= & \frac{1}{2}\int \left[ \Psi ^{\dagger }\left( t,\mathbf{r}%
\right) ,\left( -\mathrm{i}\nabla \right) \Psi \left( t,\mathbf{r}\right) %
\right] _{-}d\mathbf{r},
\notag
\\
\hat{\Xi}=& \frac{1}{2}\int \left[ \Psi
^{\dagger }\left( t,\mathbf{r}\right) ,\Xi \Psi \left( t,\mathbf{r}\right) %
\right] _{-}d\mathbf{r},  \label{3.4}
\end{align}%
where the $c$ number (generally infinite) term $H_{0}\left( t\right) $\
corresponds to the energy of vacuum fluctuations. A definition of the
corresponding vacuum is discussed just below.

Let us suppose that the effective potential $g\left( t\right) $ is constant
for $t<t_{1}$ and for $t>t_{2}$. Therefore initial (at $t<t_{1}$) and final
(at $t>t_{2}$) vacua are vacuum states of $\mathrm{in}$- and $\mathrm{out}$-
particles which correspond to the constant effective potentials $g\left(
t_{1}\right) =g_{1}$ and $g\left( t_{2}\right) =g_{2}$, respectively. During
the time interval $t_{2}$ $-t_{1}$ $=T$, the neutrino field interacts with
the time-dependent effective potential $g\left( t\right) $. The initial and
final vacua do not coincide because of the difference in the initial and
final constant values $g_{1}$ and $g_{2}$. Then we construct independently
both initial and final Fock spaces in the Heisenberg representation. We
introduce an initial set of creation and annihilation operators $%
a_{n}^{\dagger }($\textrm{in}$)$, $a_{n}($\textrm{in}$)$ of \textrm{in}%
-particles (neutrinos), and operators $b_{n}^{\dagger }($\textrm{in}$)$, $%
b_{n}($\textrm{in}$)$ of \textrm{in}-antiparticles (antineutrinos), the
corresponding \textrm{in}-vacuum being $|0,$\textrm{in}$\rangle $, and a
final set of creation and annihilation operators $a_{n}^{\dagger }($\textrm{%
out}$)$, $a_{n}($\textrm{out}$)$ of \textrm{out}-neutrinos and operators $%
b_{n}^{\dagger }($\textrm{out}$)$, $b_{n}($\textrm{out}$)$ of \textrm{out}%
-antineutrinos, the corresponding \textrm{out}-vacuum being $|0,$\textrm{out}%
$\rangle $.

Thus for any quantum number $n$, we have%
\begin{align}
& a_{n}(\mathrm{in})|0,\mathrm{in}\rangle =b_{n}(\mathrm{in})|0,\mathrm{in}%
\rangle =0,  \notag \\
& a_{n}(\mathrm{out})|0,\mathrm{out}\rangle =b_{n}(\mathrm{out})|0,\mathrm{%
out}\rangle =0.  \label{3.5}
\end{align}%
In both cases, by $n=\left( \mathbf{p},\sigma \right) $ we denote complete
sets of quantum numbers that describe both $\mathrm{in}$- and $\mathrm{out}$%
- particles and antiparticles. The $\mathrm{in}$-operators obey the
canonical anticommutation relations,%
\begin{equation}
\lbrack a_{n}(\mathrm{in}),a_{n^{\prime }}^{\dagger }(\mathrm{in}%
)]_{+}=[b_{n}(\mathrm{in}),b_{n^{\prime }}^{\dagger }(\mathrm{in}%
)]_{+}=\delta _{n,n^{\prime }}.  \label{3.6a}
\end{equation}%
All other anticommutators between the \textrm{in}-operators are equal to
zero. The $\mathrm{out}$-operators obey similar anticommutation relations, 
\begin{align}
\lbrack a_{n}(\mathrm{out}),a_{n^{\prime }}^{\dagger }(\mathrm{out}%
)]_{+}=&[b_{n}(\mathrm{out}),b_{n^{\prime }}^{\dagger }(\mathrm{out}%
)]_{+}
\notag
\\
&=\delta _{n,n^{\prime }},  \label{3.6b}
\end{align}%
and all other anticommutators between the \textrm{out}-operators also are
equal to zero.

The above $\mathrm{in}$- and $\mathrm{out}$-operators are defined by two
decompositions of the quantum Dirac field $\Psi (X)$ in the exact solutions
of the Dirac equation, 
\begin{align}
\Psi (X) = & \sum_{n}\left[ a_{n}(\mathrm{in})\;_{+}\psi
_{n}(X)+b_{n}^{\dagger }(\mathrm{in})\;_{-}\psi _{n}(X)\right]  \notag \\
& =\sum_{n}\big[ a_{n}(\mathrm{out})\;^{+}\psi _{n}\left( X\right)
\notag
\\
&+b_{n}^{\dagger }(\mathrm{out})\;^{-}\psi _{n}\left( X\right) \big] .
\label{in-out-repr}
\end{align}%
We see that the $\mathrm{in}$-operators are associated with a complete
orthonormal set of solutions $\left\{ _{\zeta }\psi _{n}(X)\right\} $ (in
the following we shall call it the $\mathrm{in}$-set) of Eq.~(\ref%
{eq:weDirHamf}) with the effective potential $g\left( t\right) $, where $%
\zeta =+$ stays for neutrinos and $\zeta =-$ for antineutrinos. Their
asymptotics at $t<t_{1}$ are wave functions of free particles in the
presence of a constant effective potential $g_{1}$ and can be classified as
neutrino and antineutrino wave functions. The $\mathrm{out}$-operators are
associated with another complete orthonormal $\mathrm{out}$-set of solutions 
$\left\{ ^{\zeta }\psi _{n}\left( X\right) \right\} $ of Eq.~(\ref%
{eq:weDirHamf}). Their asymptotics at $t>t_{2}$ are wave functions of free
particles in the presence of a constant effective potential $g_{2}$ and can
be classified as neutrino and antineutrino wave functions. The functions $%
_{\zeta }\psi _{n}(X)$ are eigenvectors of the one particle Dirac
Hamiltonian $H(t)$ at $t=t_{1}$, 
\begin{equation}
H(t_{1})_{\zeta }\psi _{n}(t_{1},\mathbf{x})=\zeta \mathcal{E}_{1}\;_{\zeta
}\psi _{n}(t_{1},\mathbf{x})\,,  \label{3.7}
\end{equation}%
where $\mathcal{E}_{1}$\ are the kinetic energies of $\mathrm{in}$-particles
(neutrino or antineutrino) in a state specified by a complete set of quantum
numbers $n$. The $\mathrm{out}$-particles (neutrino or antineutrino) are
associated with a complete $\mathrm{out}$-set of solutions $\left\{ ^{\zeta
}\psi _{n}\left( X\right) \right\} $ of the Dirac equation with the
asymptotics $^{\zeta }\psi _{n}(t_{2},\mathbf{x})$ at $t_{2}$ being
eigenvectors of the one particle Dirac Hamiltonian at $t_{2}$, namely, 
\begin{equation}
H(t_{2})\,^{\zeta }\psi _{n}(t_{2},\mathbf{x})=\zeta \mathcal{E}%
_{2}\,^{\zeta }\psi _{n}(t_{2},\mathbf{x})\,,  \label{3.8}
\end{equation}%
where $\mathcal{E}_{2}$ are the kinetic energies of $\mathrm{out}$-particles
in a state specified by a complete set of quantum numbers $n$.

One can find that for in- and out-sets, the following dispersion relations
and the orthonormality conditions hold: 
\begin{eqnarray}
&&\mathcal{E}_{1,2}=\sqrt{m^{2}+\left( p-\sigma \frac{g_{1,2}}{2}\right) ^{2}%
},  \notag \\
&&\left( _{\zeta }\psi _{n},_{\zeta ^{\prime }}\psi _{n^{\prime }}\right)
=\delta _{\zeta ,\zeta ^{\prime }}\delta _{\sigma \sigma ^{\prime }}\delta
^{\left( 3\right) }(\mathbf{p}-\mathbf{p}^{\prime }),
\notag 
\\
&&\left( ^{\zeta
}\psi _{n},^{\zeta ^{\prime }}\psi _{n^{\prime }}\right) =\delta _{\zeta
,\zeta ^{\prime }}\delta _{\sigma \sigma ^{\prime }}\delta ^{\left( 3\right)
}(\mathbf{p}-\mathbf{p}^{\prime }).  \label{3.10}
\end{eqnarray}

It should be noted, that in the following we will use the standard volume
regularization: $\delta (\mathbf{p}-\mathbf{p}^{\prime })\rightarrow \delta
_{\mathbf{p},\mathbf{p}^{\prime }}$ and $\delta _{n,n^{\prime }}=\delta
_{\sigma \sigma ^{\prime }}\delta _{\mathbf{p},\mathbf{p}^{\prime }}$.
Accounting for the orthonormality relations in Eq.~(\ref{3.10}) and the
completeness of the in- and out- sets, one can demonstrate that
anticommutation relations in Eqs.~(\ref{3.6a}) and~(\ref{3.6b}) for the
introduced creation and annihilation \textrm{in}- or \textrm{out-}operators
match with equal time anticommutation relations for the Heisenberg operators
in Eq.~(\ref{3.2}).

Being expressed in terms of the creation and annihilation operators, the
operators of physical quantities given by Eqs. (\ref{3.3}) and~(\ref{3.4})
take the form 
\begin{align}
\hat{H}\left( t_{1}\right) =&\sum_{n}\mathcal{E}_{1}\left[ a_{n}^{\dagger
}\left( \mathrm{in}\right) a_{n}\left( \mathrm{in}\right) +b_{n}^{\dagger
}\left( \mathrm{in}\right) b_{n}\left( \mathrm{in}\right) \right],
\notag
\\
H_{0}\left( t_{1}\right) = & \sum_{n}\mathcal{E}_{1},  \notag \\
\hat{H}\left( t_{2}\right) = & \sum_{n}\mathcal{E}_{2}\left[ a_{n}^{\dagger
}\left( \mathrm{out}\right) a_{n}\left( \mathrm{out}\right) +b_{n}^{\dagger
}\left( \mathrm{out}\right) b_{n}\left( \mathrm{out}\right) \right],
\notag
\\
H_{0}\left( t_{2}\right) =&\sum_{n}\mathcal{E}_{2},  \notag \\
\mathbf{\hat{p}}=&\sum_{n}\mathbf{p}\left[ a_{n}^{\dagger }\left( \mathrm{in%
}\right) a_{n}\left( \mathrm{in}\right) -b_{n}^{\dagger }\left( \mathrm{in}%
\right) b_{n}\left( \mathrm{in}\right) \right] 
\notag
\\
& =\sum_{n}\mathbf{p}\left[
a_{n}^{\dagger }\left( \mathrm{out}\right) a_{n}\left( \mathrm{out}\right)
-b_{n}^{\dagger }\left( \mathrm{out}\right) b_{n}\left( \mathrm{out}\right) %
\right] ,  \notag \\
\hat{\Xi}=&\sum_{n}\sigma \left[ a_{n}^{\dagger }\left( \mathrm{in}\right)
a_{n}\left( \mathrm{in}\right) -b_{n}^{\dagger }\left( \mathrm{in}\right)
b_{n}\left( \mathrm{in}\right) \right] 
\notag
\\
&=\sum_{n}\sigma \big[ a_{n}^{\dagger
}\left( \mathrm{out}\right) a_{n}\left( \mathrm{out}\right) 
\notag
\\
&-b_{n}^{\dagger
}\left( \mathrm{out}\right) b_{n}\left( \mathrm{out}\right) \big] .
\label{3.11}
\end{align}

We see that the creation and annihilation operators diagonalize the kinetic
energy operators $\hat{H}\left( t_{1}\right) $ and $\hat{H}\left(
t_{2}\right) $, which are positive defined. It confirms the interpretation
of the operators $a_{n}^{\dagger }\left( \mathrm{in}\right) $, $a_{n}\left( 
\mathrm{in}\right) $, $a_{n}^{\dagger }\left( \mathrm{out}\right) $, and $%
a_{n}\left( \mathrm{out}\right) $ as well as $b_{n}^{\dagger }\left( \mathrm{%
in}\right) $, $b_{n}\left( \mathrm{in}\right) $, $b_{n}^{\dagger }\left( 
\mathrm{out}\right) $, and $b_{n}\left( \mathrm{out}\right) $ as describing
a neutrino and an antineutrino at at $t=t_{1}$ and $t=t_{2}$.

As was already mentioned above, the operators $\mathbf{\hat{p}}$ and $\hat{%
\Xi}$ are the integrals of motion and are diagonal in both $\mathrm{in}$-
and $\mathrm{out}$-particle operators. Using the representations in Eq.~(\ref%
{3.11}), one can establish relations between quantum numbers $\mathbf{p}$, $%
\sigma $ and corresponding physical quantities. Namely, the physical
momentum of in- and out- neutrino is $\mathbf{p}_\mathrm{ph}=\mathbf{p}$ and
the physical helicity is $\sigma _\mathrm{ph}=\sigma $, whereas $\mathbf{p}_%
\mathrm{ph}=-\mathbf{p}$ and $\sigma _\mathrm{ph}=-\sigma $ for \textrm{in}-
and out- antineutrino. The one-particle definition of the physical helicity
operator is $\Xi _{\mathbf{p}}^\mathrm{ph}=\frac{\mathbf{p}_\mathrm{ph}%
\boldsymbol{\Sigma }}{p_\mathrm{ph}} $ for states of both neutrinos and
antineutrinos with a given momenta. It is consistent with the above given
physical interpretation of the quantum numbers $\mathbf{p}$ and $\sigma $ if
one takes into account that $\Xi _{\mathbf{p}}^\mathrm{ph}=\mathbf{p}%
\boldsymbol{\Sigma /}p$ for neutrino, whereas $\Xi _{\mathbf{p}}^\mathrm{ph}%
=-\mathbf{p}\boldsymbol{\Sigma /}p$ for antineutrino.

Further, we will see that neutrinos and antineutrinos created or annihilated
from/to the vacuum have the same quantum numbers $\mathbf{p}$ and $\sigma $
due to conservation low. This means that neutrinos and antineutrinos are
produced or annihilated with opposite physical momenta and helicities. This
matches with the interpretation given above in Sec.~\ref{SS3.1}.

$\mathrm{In}$- and $\mathrm{out}$-solutions with given quantum numbers $n$
are related by linear transformations of the form 
\begin{align}
^{\zeta }\psi _{n}\left( X\right) & =G(_{+}\mid ^{\zeta })\,_{+}\psi
_{n}\left( X\right) +G(_{-}\mid ^{\zeta })\,_{-}\psi _{n}\left( X\right) \,,
\notag \\
_{\zeta }\psi _{n}\left( X\right) & =G\left( ^{+}|_{\zeta }\right)
\,^{+}\psi _{n}\left( X\right) 
\notag
\\
&+G\left( ^{-}|_{\zeta }\right) \,^{-}\psi
_{n}\left( X\right) ,  \label{3.12}
\end{align}%
where coefficients $G$ are defined via the inner products of these sets, 
\begin{align}
\left( _{\zeta }\psi _{n^{\prime }},^{\zeta ^{\prime }}\psi _{n}\right)
= & \delta _{n,n^{\prime }}G\left( {}_{\zeta }|{}^{\zeta ^{\prime }}\right),
\notag
\\
G\left( ^{\zeta ^{\prime }}|_{\zeta }\right) = & G\left( _{\zeta }|^{\zeta
^{\prime }}\right) ^{\ast }.  \label{3.13}
\end{align}%
These coefficients satisfy the unitarity relations 
\begin{align}
& G\left( ^{\zeta }|_{+}\right) G\left( {}_{+}|{}^{\zeta }\right) +G\left(
^{\zeta }|_{-}\right) G\left( {}_{-}|{}^{\zeta }\right) =1\,,  \notag \\
& G\left( {}_{\zeta }|{}^{+}\right) G\left( ^{+}|_{\zeta }\right) +G\left(
{}_{\zeta }|{}^{-}\right) G\left( ^{-}|_{\zeta }\right) =1\,,  \notag \\
& G\left( {}_{+}|{}^{+}\right) G\left( ^{+}|_{-}\right) +G\left(
{}_{+}|{}^{-}\right) G\left( ^{-}|_{-}\right) =0\,,  \notag \\
& G\left( ^{+}|_{+}\right) G\left( {}_{+}|{}^{-}\right) +G\left(
^{+}|_{-}\right) G\left( {}_{-}|{}^{-}\right) =0\,,  \label{3.14}
\end{align}%
which follow from the orthonormalization and completeness relations for the
corresponding solutions. It is known that all the coefficients can be
expressed in terms of two of them, e.g., of $G\left( _{+}\left\vert
^{+}\right. \right) $ and $G\left( _{-}\left\vert ^{+}\right. \right) $.
However, even these coefficients are not completely independent, 
\begin{equation}
\left\vert G\left( _{-}\left\vert ^{+}\right. \right) \right\vert
^{2}+\left\vert G\left( _{+}\left\vert ^{+}\right. \right) \right\vert
^{2}=1.  \label{3.15}
\end{equation}

A linear canonical transformation (Bogolyubov transformation) between $%
\mathrm{in}$- and $\mathrm{out}$- operators which can be derived from Eq.~(%
\ref{in-out-repr}) has the form 
\begin{align}
a_{n}\left( \mathrm{out}\right) & =G\left( ^{+}|_{+}\right) a_{n}(\mathrm{in}%
)+G\left( ^{+}|_{-}\right) b_{n}^{\dagger }(\mathrm{in}),  \notag \\
b_{n}^{\dagger }\left( \mathrm{out}\right) & =G\left( ^{-}|_{+}\right) a_{n}(%
\mathrm{in})+G\left( ^{-}|_{-}\right) b_{n}^{\dagger }(\mathrm{in}).
\label{3.16}
\end{align}

All the information about neutrino and antineutrino creation, annihilation,
and scattering in a background matter can be extracted from the coefficients 
$G\left( {}_{\zeta }|{}^{\zeta ^{\prime }}\right) $. For example, using Eq.~(%
\ref{3.16}), we find the differential mean number $N_{n}$ of neutrino or
antineutrino created (which are also equal to the mean number of $\nu \bar{%
\nu}$ pairs created)\textbf{\ }from the \textrm{in}-vacuum with a given
momentum $\mathbf{p}$ and spin projection $\sigma $ is 
\begin{equation}
N_{n}=\langle 0,\mathrm{in}|a_{n}^{\dagger }(\mathrm{out})a_{n}(\mathrm{out}%
)|0,\mathrm{in}\rangle =\left\vert G\left( {}_{-}|{}^{+}\right) \right\vert
^{2}.  \label{3.17}
\end{equation}%
The total number $\mathcal{N}_{\sigma }$ of created $\nu \bar{\nu}$ pairs
with a given $\sigma $ is the sum over all the momenta,%
\begin{equation}
\mathcal{N}_{\sigma }=\sum_{\mathbf{p}}N_{n}=\frac{V}{\left( 2\pi \right)
^{3}}\int N_{n}d\mathbf{p}.  \label{TN}
\end{equation}%
The probability of the neutrino scattering $P(+|+)_{n,n^{\prime }}$ and the
probability of a pair creation $P(-+|0)_{n,n^{\prime }}$ are, respectively 
\begin{align}
P(+|+)_{n,n^{\prime }}=&|\langle 0,\mathrm{out}|a_{n}(\mathrm{out}%
)a_{n^{\prime }}^{\dagger }(\mathrm{in})|0,\mathrm{in}\rangle |^{2}
\notag
\\
&=\delta
_{n,n^{\prime }}\frac{1}{1-N_{n}}P_{v}\;,  \notag \\
P(-+|0)_{n,n^{\prime }}=&|\langle 0,\mathrm{out}|b_{n}(\mathrm{out}%
)a_{n^{\prime }}(\mathrm{out})|0,\mathrm{in}\rangle |^{2}
\notag
\\
&=\delta
_{n,n^{\prime }}\frac{N_{n}}{1-N_{n}}P_{v}\;.  \label{3.18}
\end{align}%
The probability for the neutrino vacuum to remain a vacuum reads%
\begin{equation}
P_{v}=\left\vert \left\langle 0,\mathrm{out}|0,\mathrm{in}\right\rangle
\right\vert ^{2}=\exp \left\{ \sum_{\sigma ,\mathbf{p}}\ln \left(
1-N_{n}\right) \right\} .  \label{3.19}
\end{equation}%
The probabilities for an antineutrino scattering and a $\nu \bar{\nu}$ pair
annihilation are given by the same expressions $P(+|+)$ and $P(-+|0)$,
respectively.

In the general case, states of the system under consideration at the final
time instant contain particles and antiparticles due to the $\nu \bar{\nu}$
pair creation from the vacuum and due to the possible existence of some
particles and antiparticles (we call them initial in what follows) in the
initial state of the system. It was found in Ref.~\cite{GavGT06} that the
following relation holds true: 
\begin{align}
\aleph _{n}^{(\zeta )}\left( \mathrm{out}\right) = & \left( 1-N_{n}\right)
\aleph _{m}^{(\zeta )}(\mathrm{in})
\notag
\\
&+N_{n}\left[ 1-\aleph _{n}^{(-\zeta )}(%
\mathrm{in})\right] ,  \label{3.20}
\end{align}%
where $\aleph _{n}^{(\zeta )}($\textrm{in}$)$ and $\aleph _{n}^{(\zeta
)}\left( \mathrm{out}\right) $ are initial and the final differential mean
numbers of particles ($\zeta =+$) and antiparticles ($\zeta =-$). Here $%
N_{n} $ is given by Eq.~(\ref{3.17}). Thus, if the initial state differs
from the vacuum, the differential mean numbers of neutrinos or antineutrinos
created by the effective potential $g\left( t\right) $ are given by the
difference $\Delta \aleph _{n}^{(\zeta )}=\aleph _{n}^{(\zeta )}\left( 
\mathrm{out}\right) -\aleph _{n}^{(\zeta )}($\textrm{in}$)$.

Using Eq.~(\ref{3.20}), we obtain that 
\begin{eqnarray}
&&\Delta \aleph _{n}^{(+)}=\Delta \aleph _{n}^{(-)}=\Delta \aleph _{n}\,, 
\notag \\
&&\Delta \aleph _{n}=N_{n}\left[ 1-\left( \aleph _{n}^{(+)}(\mathrm{in}%
)+\aleph _{n}^{(-)}(\mathrm{in})\right) \right] \,.  \label{3.21}
\end{eqnarray}%
Even if $N_{n}\neq 0,$ no creation of $\nu \bar{\nu}$-pairs with quantum
numbers $n$ occurs provided that $N_{n}^{(+)}($\textrm{in}$)+N_{n}^{(-)}($%
\textrm{in}$)=1$. It happens because of the Pauli blocking when both
particle and antiparticle are involved. The $\nu \bar{\nu}$ pairs creation
takes place if $N_{n}^{(+)}($\textrm{in}$)+N_{n}^{(-)}($\textrm{in}$)<1$.
The annihilation of $\nu \bar{\nu}$ pairs\ is possible if $N_{n}^{(+)}($%
\textrm{in}$)+N_{n}^{(-)}($\textrm{in}$)>1$.

\section{Neutrino creation by a slowly varying effective potential\label{S4}}

In this section we study creation of $\nu \bar{\nu}$ pairs of various
neutrino flavors by a background matter with a linearly growing effective
potential. We consider the so-called strong field case, when the difference $%
\left\vert g(t_{\mathrm{out}})\right\vert -\left\vert g(t_{\mathrm{in}%
})\right\vert $ between the initial and final potential is greater then the
neutrino mass $m$. In this sense, one can say that an effective potential $%
g\left( t\right) $ is slowly varying.

To find all necessary ingredients for calculating the particle-creation
effect, we first represent solutions $\tilde{\psi}\left( X\right) $ of Eq. (%
\ref{eq:weDirHamf}) in the following form,%
\begin{equation}
\psi _{n}\left( X\right) =\left[ \mathrm{i}\partial _{0}+H\left( t\right) %
\right] \varphi _{n,\chi }\left( t\right) e^{\mathrm{i}\mathbf{pr}}U_{\sigma
,\chi },  \label{4.1}
\end{equation}%
where $\varphi _{n,\chi }\left( t\right) $ are time-dependent scalar
functions that satisfy the equation 
\begin{align}
&\left[ \frac{\mathrm{d}^{2}}{\mathrm{d}t^{2}}+\left( \sigma p-\frac{g\left(
t\right) }{2}\right) ^{2}+\frac{\mathrm{i}}{2}\chi \partial _{t}g\left(
t\right) +m^{2}\right]
\notag
\\
&\times
\varphi _{n,\chi }\left( t\right) =0,  \label{4.3}
\end{align}%
whereas constant spinors $U_{\sigma ,\chi }$ satisfy the equations 
\begin{align}
\frac{\mathbf{p}\boldsymbol{\Sigma }}{p}U_{\sigma ,\chi }=&\sigma U_{\sigma
,\chi },\quad \sigma =\pm 1;
\notag 
\\
\gamma ^{5}U_{\sigma ,\chi }=&\chi
U_{\sigma ,\chi },\quad \chi =\pm 1.  \label{4.2}
\end{align}

Note that $\gamma ^{5}$ does not commute with the projection operator in the
representation given in Eq.~(\ref{4.1}). Therefore solutions $\psi
_{n}\left( X\right) $ that correspond to different spinors $U_{\sigma ,+1}$
and $U_{\sigma ,-1}$ are linear dependent. Then one can choose, for example,
either $\chi =+1$ or $\chi =-1$.

Using Eq.~(\ref{4.1}), we express the inner product (\ref{1.6}) of two
arbitrary solutions $\tilde{\psi}_{n}\left( X\right) $ and $\tilde{\psi}%
_{n}^{\prime }\left( X\right) $ as follows%
\begin{align}
\left( \psi _{n},\psi _{n^{\prime }}^{\prime }\right) =&\delta
_{n,n^{\prime }}VJ,  \notag \\
J=&U_{\sigma ,\chi }^{\dag }\varphi _{n,\chi }^{\ast }\left( t\right)
\left( -\mathrm{i}\overleftarrow{\partial }_{0}+\mathrm{i}\partial
_{0}\right)
\notag
\\
&\times
\left[ \mathrm{i}\partial _{0}+\chi \left( p\sigma -\frac{%
g\left( t\right) }{2}\right) +m\gamma ^{0}\right]
\notag
\\
&\times
\varphi _{n,\chi }^{\prime
}\left( t\right) U_{\sigma ,\chi }.  \label{4.4}
\end{align}%
Then, we obtain the quantity $J$ in the following form 
\begin{align}
J=&\delta _{n,n^{\prime }}\varphi _{n,\chi }^{\ast }\left( t\right) \left( -%
\mathrm{i}\overleftarrow{\partial }_{0}+\mathrm{i}\partial _{0}\right)
\notag
\\
&\times
\left[
\mathrm{i}\partial _{0}+\chi \left( p\sigma -\frac{g\left( t\right) }{2}%
\right) \right] \varphi _{n,\chi }^{\prime }\left( t\right) .  \label{4.5}
\end{align}

Setting $t=t_{1}$ and $t=t_{2}$ in Eqs.~(\ref{3.7}), (\ref{3.8}), and (\ref%
{4.4}), one gets that particle and antiparticle degrees of freedom are
simultaneously orthogonal: $\left( _{+}\psi _{n},_{-}\psi _{n^{\prime
}}\right) =\left( ^{+}\psi _{n},^{-}\psi _{n^{\prime }}\right) =0$. We see
that here it is enough to know only scalar functions in Eq.~(\ref{4.1}). The
same holds true for the calculation of all other necessary quantities.

Now we consider the case of a slowly varying effective potential supposing
that $g\left( t\right) $ is a linear function in a rather big time interval $%
T=t_{2}-t_{1}$. Namely, we are going to consider the following time
dependence of effective potential, 
\begin{equation}
g(t)=%
\begin{cases}
g_{1}, & t<t_{1}, \\ 
b-at, & t_{1}\leq t\leq t_{2}, \\ 
g_{2}, & t>t_{2},%
\end{cases}
\label{4.6}
\end{equation}%
where $g(t_{1})=g_{1}$ and $g(t_{2})=g_{2}$ are constant values and 
\begin{equation}
a=-\frac{g_{2}-g_{1}}{t_{2}-t_{1}}\neq 0,\quad b=\frac{g_{1}t_{2}-g_{2}t_{1}%
}{t_{2}-t_{1}}.  \label{eq:abdef}
\end{equation}%
We shall study the $\nu \bar{\nu}$ pairs creation due to the compression
before the hydrodynamic bounce which happens during $0.10\,\text{s}\lesssim
t\lesssim 0.11\,\text{s}$ ($t=0$ corresponds to the beginning of the
collapse) and during the neutronization of PNS which occurs during $0.11\,%
\text{s}\lesssim t\lesssim 0.12\,\text{s}$ (for the details see Ref.~\cite%
{Jan07} and Sec.~\ref{S5}). If we study the pairs creation \ due to the
matter compression in the PNS core, using Eqs.~(\ref{eq:gexpl}) and (\ref%
{1.4}), we obtain, for example, that $g_{1}=g(t_{\mathrm{in}})\approx 0$ for
all the neutrino flavors, whereas $g_{2}=g(t_{\mathrm{out}})=0$ for $\nu
_{e} $ and $g_{2}=g(t_{\mathrm{out}})<0$ for $\nu _{\mu ,\tau }$. If we
examine the vacuum instability in the neutronization of PNS that occurs%
\textbf{\ }outside the core, then $g_{2}=g(t_{\mathrm{out}})<0$ for all the
neutrino flavors. However $g_{1}=g(t_{\mathrm{in}})>0$ for $\nu _{e}$ and $%
g_{1}=g(t_{\mathrm{in}})<0$ for $\nu _{\mu }$ and $\nu _{\tau }$. We can
always choose $t_{1,2}$ to have $b=0$ in Eq.~(\ref{eq:abdef}). The model
with the external field $g\left( t\right) $ given by Eq.~(\ref{4.6}) is
technically similar to the QED model with the $T$-constant external electric
field studied in Ref.~\cite{GavG96} and can be treated similarly.

First of all, we consider solutions in Eq.~(\ref{4.1}) at $t<t_{1}$ and $%
t>t_{2}$ corresponding to the constant effective potential $g_{1}$ or $g_{2}$%
, respectively. We present such solutions in the following normalized form 
\begin{eqnarray}
&&_{\zeta }\psi _{n}\left( X\right) =\left[ \mathrm{i}\partial _{0}+H\left(
t\right) \right] \;_{\zeta }\varphi _{n,\chi }\left( t\right) e^{\mathrm{i}%
\mathbf{pr}}U_{\sigma ,\chi },
\notag
\\
&&_{\zeta }\varphi _{n,\chi }\left( t\right)
=\;C_{1}^{\zeta }\exp \left[ -i\zeta \mathcal{E}_{1}(t-t_{1})\right] ,%
\mathrm{\;}t<t_{1},  \notag \\
&&^{\zeta }\psi _{n}\left( X\right) =\left[ \mathrm{i}\partial _{0}+H\left(
t\right) \right] \;^{\zeta }\varphi _{n,\chi }\left( t\right) e^{\mathrm{i}%
\mathbf{pr}}U_{\sigma ,\chi },
\notag
\\
&&^{\zeta }\varphi _{n,\chi }\left( t\right)
=\;C_{2}^{\zeta }\exp \left[ -i\zeta \mathcal{E}_{2}(t-t_{2})\right] ,%
\mathrm{\;}t>t_{2},  \notag \\
&&C_{1,2}^{\zeta }=(2V\mathcal{E}_{1,2})^{-1/2}\left\vert \mathcal{E}%
_{1,2}-\zeta \chi \left( \frac{g_{1,2}}{2}-\sigma p\right) \right\vert
^{-1/2} ,  \label{4.7}
\end{eqnarray}%
where neutrino and antineutrino states are identified according to the
kinetic energy signs in Eqs.~(\ref{3.7}) and (\ref{3.8}). Normalization
factors $C_{1,2}^{\zeta }$ are calculated in accordance with Eqs.~(\ref{4.4}%
) and~(\ref{4.5}).

Using representations in Eq.~(\ref{4.7}), we can reproduce solutions of the
Dirac equation obtained in Sec.~\ref{SS3.1}. Indeed, let us write%
\begin{equation}
\varphi _{n,\chi }\sim \exp \left( \mp \mathrm{i}\mathcal{E}_{1,2}t+\mathrm{i%
}\mathbf{pr}\right) ,\;U_{\sigma ,\chi }\sim \left( 
\begin{array}{c}
w_{\sigma } \\ 
\chi w_{\sigma }%
\end{array}%
\right) .  \label{4.7a}
\end{equation}%
Using the explicit form of $\gamma $ matrices in Eq.~(\ref{eq:Dirmat}) one
can verify that Eq.~(\ref{4.2}) holds true. Then we see that, for $\chi =+1$%
, the corresponding neutrino wave functions $\psi _{n}(X)\exp \left( -%
\mathrm{i}tg_{1,2}/2\right) $ coincide with the function given by Eqs.~(\ref%
{psipmrqm}) and~(\ref{eq:basspinDir}) up to constant factors. Thus, neutrino
wave functions considered in Sec.~\ref{SS3.1} are consistent with wave
functions that are obtained for time-dependent effective potentials (see
also Ref.~\cite{StuTer05}).

Now, we consider solutions (\ref{4.1}) at $t_{1}\leq t\leq t_{2}$. In this
time region, the functions $\varphi _{n,\chi }\left( t\right) $ satisfy the
following equation:%
\begin{equation}
\left[ \frac{\mathrm{d}^{2}}{\mathrm{d}\xi ^{2}}+\xi ^{2}-\mathrm{i}\chi 
\mathrm{sgn}(a)+\lambda \right] \varphi _{n,\chi }\left( t\right) =0,
\label{eq:feq}
\end{equation}%
where $\lambda =2m^{2}/\left\vert a\right\vert $ and 
\begin{equation}
\xi =\sqrt{\frac{2}{\left\vert a\right\vert }}\left( \frac{a}{2}t-\frac{b}{2}%
+\sigma p\right) \mathrm{sgn}(a).  \label{eq:xidef}
\end{equation}

For $\chi \mathrm{sgn}(a)=+1$, one can see that two independent solutions of
Eq.~(\ref{eq:feq}) are $D_{\rho }[(1-\mathrm{i})\xi ]$ and $D_{-1-\rho }[(1+%
\mathrm{i})\xi ],$ where $D_{\rho }(\xi )$ is Weber parabolic cylinder
function (WPCF) and $\rho =\mathrm{i}\lambda /2$. It is known that these
solutions form a complete set. Some useful properties of these solutions are
summarized in Appendix~\ref{App} and will be used in what follows.

To obtain the coefficient $G\left( _{-}\left\vert ^{+}\right. \right) $,
corresponding to the time-dependent effective potential in Eq.~(\ref{4.6}),
we use Eq.~(\ref{4.5}). Since the inner product in Eq.~(\ref{4.4}) is time
independent we can use any convenient time instant for it calculation. Let
us set $t=t_{0}<t_{1}$ in Eq.~(\ref{4.5}). Then we have to use the
corresponding functions $_{-}\varphi _{n,\chi }\left( t\right) $ from Eq.~(%
\ref{4.7}). According to Eq.~(\ref{3.12}) the function $^{+}\varphi _{n,\chi
}\left( t\right) $ for any time instant can be presented in the form
\begin{widetext} 
\begin{equation}
^{+}\varphi _{n,\chi }\left( t\right) =%
\begin{cases}
G(_{+}|^{+})_{+}\varphi _{n,\chi }\left( t\right) +G(_{-}|^{+})_{-}\varphi
_{n,\chi }\left( t\right) , & t<t_{1}, \\ 
C_{2}^{+}\left( d_{1}D_{\rho }[(1-i)\xi ]+d_{2}D_{-1-\rho }[(1+i)\xi
]\right) , & t_{1}\leq t\leq t_{2}, \\ 
C_{2}^{+}\exp \left[ -i\mathcal{E}_{2}(t-t_{2})\right] , & t>t_{2}.%
\end{cases}
\label{eq:f2}
\end{equation}%
\end{widetext}
The coefficients $d_{1,2}$ will be specified below. The functions $%
^{+}\varphi _{n,\chi }\left( t\right) $ and their derivatives $\partial
_{t}\;^{+}\varphi _{n,\chi }\left( t\right) $ satisfy the following gluing
conditions: 
\begin{align}
^{+}\varphi _{n,\chi }(t_{k}-0)=&^{+}\varphi _{n,\chi }\left( t\right)
(t_{k}+0),
\notag
\\
\partial _{t}\;^{+}\varphi _{n,\chi }(t_{k}-0)=&\partial
_{t}\;^{+}\varphi _{n,\chi }\left( t\right) (t_{k}+0),
\notag
\\
k=&1,2.
\label{eq:contcond}
\end{align}%
Let us choose, for example, $\chi \mathrm{sgn}(a)=+1$. Then, at $t=t_{2},$
it follows from Eq.~(\ref{eq:contcond}) that%
\begin{equation}
d_{1,2}=\mp \frac{\mathcal{E}_{2}}{\sqrt{a}\exp [(\lambda -\mathrm{i})\pi /4]%
}f_{1,2}(t_{2}),  \label{200}
\end{equation}%
where 
\begin{align}
f_{1}(t)=& \left\{ 1-\frac{\mathrm{i}}{\sqrt{\xi ^{2}+\lambda }}\frac{%
\mathrm{d}}{\mathrm{d}\xi }\right\} D_{-1-\mathrm{i}\lambda /2}[(1+\mathrm{i}%
)\xi ],  \notag \\
f_{2}(t)=& \left\{ 1-\frac{\mathrm{i}}{\sqrt{\xi ^{2}+\lambda }}\frac{%
\mathrm{d}}{\mathrm{d}\xi }\right\} D_{\mathrm{i}\lambda /2}[(1-\mathrm{i}%
)\xi ].  \label{4.8}
\end{align}%
Finally, applying Eq.~(\ref{eq:contcond}) at $t=t_{1}$, we get $G\left(
_{-}\left\vert ^{+}\right. \right) $ in the following form: 
\begin{eqnarray}
&&G\left( _{-}\left\vert ^{+}\right. \right) =\exp [-(\lambda -\mathrm{i}%
)\pi /4]AB,
\notag
\\
&&B=\left[ f_{1}(t_{1})f_{2}(t_{2})-f_{2}(t_{1})f_{1}(t_{2})%
\right] ,  \notag \\
&&A=\left[ \frac{\sqrt{\xi _{1}^{2}+\lambda }\sqrt{\xi _{2}^{2}+\lambda }%
\left( \sqrt{\xi _{1}^{2}+\lambda }-\xi _{1}\right) }{8\sqrt{\xi
_{2}^{2}+\lambda }+\xi _{2}}\right] ^{1/2},  \label{4.9}
\end{eqnarray}%
where%
\begin{equation}
\xi _{1,2}=\left. \xi \right\vert _{t=t_{1,2}}=\sqrt{\frac{2}{\left\vert
a\right\vert }}\left( \sigma p-\frac{g_{1,2}}{2}\right) \mathrm{sgn}(a),
\label{4.9a}
\end{equation}

According to Eq.~(\ref{3.17}), the differential mean numbers of the $\nu 
\bar{\nu}$ pairs created by the effective potential Eq.~(\ref{4.6}) are 
\begin{equation}
N_{n}=\left\vert G\left( {}_{-}|{}^{+}\right) \right\vert ^{2}=e^{-\pi
\lambda /2}A^{2}\left\vert B\right\vert ^{2}.  \label{4.10}
\end{equation}%
They depend only on the values $\xi _{1,2}$ for a given $\lambda $. Similar
expressions were obtained in Ref.~\cite{GavG96} in the problem of particle
creation by a quasiconstant uniform electric field.

We are interested in the case of a slowly varying strong effective potential 
$g\left( t\right) $, that satisfies the condition%
\begin{align}
\left\vert g_{2}-g_{1}\right\vert \left\vert a\right\vert ^{-1/2}=&\left[
\left\vert g_{2}-g_{1}\right\vert \left( t_{2}-t_{1}\right) \right]
^{1/2}
\notag
\\
&\gg K\gg \max \left\{ 1,\lambda \right\} ,  \label{4.11}
\end{align}%
where $K$ is a given number. The case when both $\left\vert \xi
_{1}\right\vert $ and $\left\vert \xi _{2}\right\vert $ are sufficiently
large, 
\begin{equation}
\left\vert \xi _{1,2}\right\vert \geq K\gg \max \left\{ 1,\lambda \right\} ,
\label{4.12}
\end{equation}%
is only possible when signs of $\xi _{1}$ and $\xi _{2}$ are opposite. In
this case, using asymptotic expansions of WPCF, we obtain (see details in
Appendix \ref{App}) that 
\begin{equation}
N_{n}=e^{-\pi \lambda }\left[ 1+O\left( \left\vert \xi _{1}\right\vert
^{-3}\right) +O\left( \left\vert \xi _{2}\right\vert ^{-3}\right) \right] .
\label{4.13}
\end{equation}%
Consequently, the quantity (\ref{4.13}) is almost constant over the wide
range of momenta if Eq.~(\ref{4.12}) holds true. For the case of
sufficiently big momenta, when $\xi _{1}\approx \xi _{2}$, we find that the
quantity $N_{n}$ is very small, 
\begin{align}
N_{n}\sim & \max \left\{ \left\vert \xi _{1}\right\vert ^{-6},\left\vert \xi
_{2}\right\vert ^{-6}\right\} \;\;\mathrm{if}
\notag
\\
&\min \left\{ \left\vert \xi
_{1}\right\vert ,\left\vert \xi _{2}\right\vert \right\} \geq K.
\label{4.14}
\end{align}

\begin{figure}[tbp]
\centering\includegraphics[scale=.55]{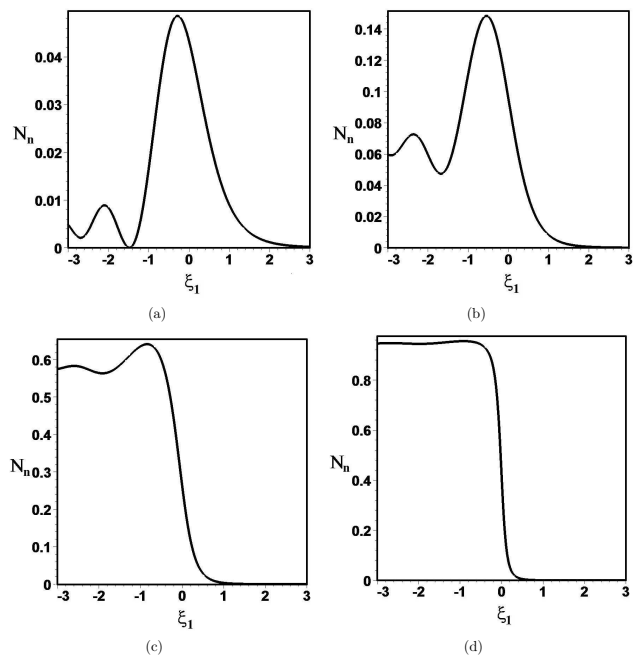}
\caption{ The dependence of the number of $N_{n}$ in Eq.~(\protect\ref%
{eq:Nnf1}) versus $\protect\xi _{1}$ for different $\protect\lambda $. The
panel (a) corresponds to $\protect\lambda =2$, the panel (b) -- to $\protect%
\lambda =1$, the panel (c) -- to $\protect\lambda =0.2$, and the panel (d)
-- to $\protect\lambda =0.02$.}
\label{fig:Nnf1}
\end{figure}

In the intermediate region the values of $\left\vert \xi _{1}\right\vert $
and $\left\vert \xi _{2}\right\vert $ are quite different. For example, when 
$\left\vert \xi _{2}\right\vert \geq K$ then $\left\vert \xi _{1}\right\vert
<K$ and vice versa. Thus, here, we cannot use any asymptotic expansion of
WPCFs to analyze the $\xi _{1}$-dependence of $N_{n}$. However, one can make
some conclusions about the contribution of this region to the integral over
the momenta in Eqs.~(\ref{TN}). Taking into account that $N_{n}$ is always
smaller than one for fermions, one can get a rough estimation%
\begin{align*}
\int_{\left\vert \xi _{1}\right\vert <K}N_{n}dp
<&\int_{\left\vert \xi
_{1}\right\vert <K}dp
\\
&\sim V\mathbf{\max }\left\{ \sqrt{\left\vert
a\right\vert }K\left\vert g_{1}\right\vert ^{2},\left( \sqrt{\left\vert
a\right\vert }K\right) ^{3}\right\} .
\end{align*}%
A more accurate estimations can be made numerically. We assume that $\xi
_{2}\geq K$ and $\left\vert \xi _{1}\right\vert <K$. Using the only
asymptotics with respect to $\xi _{2}$ given by Eq.~(\ref{a4}) and the exact
form of $f_{1}(t_{1})$ given by Eq.~(\ref{4.8}), we find that 
\begin{align}
N_{n}=&\frac{1}{4}e^{-\pi \lambda /4}\sqrt{\xi _{1}^{2}+\lambda }
\notag
\\
&\times
\left( \sqrt{%
\xi _{1}^{2}+\lambda }-\xi _{1}\right) \left\vert f_{1}(t_{1})\right\vert
^{2},  \label{eq:Nnf1}
\end{align}%
exactly in $\xi _{1}$. The dependence on $\xi _{1}$ of $N_{n}$ given by Eq.~(%
\ref{eq:Nnf1}) is made numerically for different $\lambda $ and is presented
on Fig.~\ref{fig:Nnf1}. Thus, we find that the contribution from the
intermediate region to the integral in Eq.~(\ref{TN}) is much less than that
given by a rough estimate. In particular, we show that the value $K=3$ is
sufficiently large for the problem in question.

Thus, the parameter $K$ plays the role of a sharp cutoff in the integral in
Eq.~(\ref{TN}). Finally we find that the differential mean numbers of
neutrinos or antineutrinos can be written as%
\begin{equation}
N_{n}=\left[ 
\begin{array}{l}
e^{-\pi \lambda },\ \ \mathbf{p}\in D_{\sigma } \\ 
0,\ \ \mathbf{p}\notin D_{\sigma }%
\end{array}%
\right. ,  \label{4.15a}
\end{equation}%
where%
\begin{align}
D_{\sigma }:\left\vert \xi _{1,2}\right\vert \geq & K\gg \max \left\{
1,\lambda \right\} ,
\notag
\\
&\mathrm{sgn}\left( \xi _{1}\right) =-\mathrm{sgn}%
\left( \xi _{2}\right) .  \label{4.15}
\end{align}%
We see that in the range $D_{\sigma }$ the distribution $N_{n}$ is uniform
and rotationally invariant and is completely determined by the value of $%
\lambda $.

We can conditionally consider $\lambda \lesssim 1$ as a characteristic of
the strong-field case, and $\lambda \gg 1$ as a characteristic of the
weak-field case. The effect of particle creation is negligible small in the
latter case. Here we have similar situation with the charged particle
creation by an electric field $E$ from the vacuum, where there exists
similar parameter $m^{2}/eE$ and its characteristic value $m^{2}/eE=1$
defines the Schwinger's critical field $E_{\mathrm{cr}}=m^{2}/e$.

In the following we assume that in our problem $\lambda \lesssim 1$ and
define the critical neutrino mass $m^{(\mathrm{cr})}$ from the condition $%
\lambda =1.$ Obviously, the effect of neutrino creation can be in principle
observed if there exists a kind of neutrinos with masses that are less or
comparable with such a critical mass. For the further estimations, it is
convenient, using the definition of $a$ in Eq.~(\ref{eq:abdef}), to express $%
\lambda $ as follows 
\begin{equation}
\lambda =\frac{2m^{2}\left( t_{2}-t_{1}\right) }{\left\vert
g_{2}-g_{1}\right\vert }.  \label{4.16}
\end{equation}

The total number $\mathcal{N}_{\sigma }$ of neutrino or antineutrino with a
given $\sigma $ created from vacuum is proportional to the total number of
states $\Delta _{\sigma }$ with the neutrino momenta that belong to the
range $D_{\sigma }$. Thus, we have 
\begin{equation}
\mathcal{N}_{\sigma }=e^{-\pi \lambda }\Delta _{\sigma },\;\;\Delta _{\sigma
}=\frac{V}{\left( 2\pi \right) ^{3}}\int_{D_{\sigma }}d\mathbf{p}.
\label{4.17}
\end{equation}

The logarithm of the probability for the neutrino vacuum to remain a vacuum
given by Eq.~(\ref{3.19}) is also proportional to $\Delta _{\sigma }$, 
\begin{equation}
\ln P_{v}=\ln \left( 1-e^{-\pi \lambda }\right) \left( \Delta _{+1}+\Delta
_{-1}\right) \;.  \label{4.17.1}
\end{equation}%
Note that if $e^{-\pi \lambda }\ll 1$\ then $\ln P_{v}\approx -\left( 
\mathcal{N}_{+1}+\mathcal{N}_{-1}\right) $.

The energy density of created neutrino or antineutrino with a given $\sigma $
has the form 
\begin{equation}
w_{\sigma }=\frac{e^{-\pi \lambda }}{\left( 2\pi \right) ^{3}}%
\int_{D_{\sigma }}\mathcal{E}_{2}d\mathbf{p},  \label{4.17.2}
\end{equation}%
where $\mathcal{E}_{2}$ is defined by Eq.~(\ref{3.10}). In the strong-field
case defined just above, the dependence on the cutoff $K$ can be ignored in
Eqs.~(\ref{4.17})-(\ref{4.17.2}).

Considering other models with slowly varying effective potentials that
correspond to the strong field case, cf. Ref.~\cite{GavG96}, one can verify
that effects of switching on and off do not change essentially the form of
the distribution~(\ref{4.15a}) if some conditions~similar to the one (\ref%
{4.11}) are fulfilled.

As was mentioned in Sec.~\ref{S2}, we suppose that transitions between
eigenstates that correspond to different neutrino flavors are suppressed. In
such a case, we suppose that there exist three effective masses $m_{\nu
_{e}} $, $m_{\nu _{\mu }}$, and $m_{\nu _{\tau }}$ of three active neutrino
flavors $\nu _{e}$, $\nu _{\mu }$, and $\nu _{\tau }$. Of course, all the
results obtained above for a single mass $m$ hold true for each mass $%
m=m_{\alpha }$, where $\alpha =\nu _{e,\mu ,\tau }$. Since the problem of
the neutrino masses hierarchy is still an open question~\cite{Cah13}, any
one of these masses can be critical. That is why we have to consider all the
possibilities. We denote the parameters (\ref{4.16}) by $\lambda _{e}$, $%
\lambda _{\mu }$, and $\lambda _{\tau }$ for $m_{\nu _{e}}$, $m_{\nu _{\mu
}} $, and $m_{\nu _{\tau }}$ respectively.

The difference in the effective potentials for $\nu _{e}$ and $\nu _{\mu
,\tau }$ in Eq.~(\ref{eq:gexpl}) implies the difference in the momentum
ranges of the corresponding neutrinos created at the neutronization stage.
We assume that $a>0$. Then, e.g. it results from Eq.~(\ref{1.4}) that $%
g_{1}=g(t_{\mathrm{in}})>0$ and $g_{2}=g(t_{\mathrm{out}})=-2g(t_{\mathrm{in}%
})$ for $\nu _{e}$. Using Eq. (\ref{4.15}), we find that the maximal range
of $\nu _{e}$ momenta is%
\begin{eqnarray}
D_{-1}^{e} &:&p\leq \frac{\left\vert g_{2}\right\vert }{2}-\sqrt{\frac{a}{2}}%
K\;\;\mathrm{if}\;\;\sigma =-1,  \notag \\
D_{+1}^{e} &:&p\leq \frac{g_{1}}{2}-\sqrt{\frac{a}{2}}K\;\;\mathrm{if}%
\;\;\sigma =+1.  \label{4.18}
\end{eqnarray}%
We see that it depends on the neutrino helicity.

The total number of states $\Delta _{\sigma }^{e}$ in the range given by Eq.
(\ref{4.15}) can be considered as the function of the interval $%
T=t_{2}-t_{1} $ of the effective potential variation. Note that one can take
any value of $g_{1}\geq g(t_{\mathrm{in}})$ as initial and $g_{2}\leq g(t_{%
\mathrm{out}})$ as final unless the condition (\ref{4.11}) is fulfilled for
these quantities. Then specific intervals of a pair formation can be
determined. In particular, one can find ranges of the momenta for the $\nu
_{e}$ created before the value $g(t)$ decreases to zero at some time $t_{0}$
($g(t)>0$ part) and after that ($g(t)<0$ part). In the first situation, one
has $g_{1}=g(t_{\mathrm{in}})$ and $g_{2}=0,$ while in the second, $g_{1}=0$
and $g_{2}=g(t_{\mathrm{out}})$. Then not empty ranges are 
\begin{align}
&D_{-1}^{e}\left( g(t)<0\right) :\sqrt{\frac{a}{2}}K\leq p\leq \frac{%
\left\vert g_{2}\right\vert }{2}-\sqrt{\frac{a}{2}}K,
\notag
\\
&\mathrm{for}%
\;\;g(t)<0,\;\;\mathrm{if}\;\;\sigma =-1,  \notag \\
&D_{+1}^{e}\left( g(t))>0\right) :\sqrt{\frac{a}{2}}K\leq p\leq \frac{g_{1}%
}{2}-\sqrt{\frac{a}{2}}K,
\notag
\\
&\mathrm{for}\;\;g(t)>0,\;\;\mathrm{if}%
\;\;\sigma =+1.  \label{4.19}
\end{align}%
We see that the $\nu _{e}\bar{\nu}_{e}$ pairs with $\sigma =-1$ are mainly
created when the potential $g(t)$ becomes negative, in contrast to the $\nu
_{e}\bar{\nu}_{e}$ pairs with $\sigma =+1$ that are created earlier. Using
Eqs.~(\ref{4.18}) and (\ref{4.19}), we find that the maximal kinetic energy
of created electron neutrino or antineutrino at final time instant of the
neutronization $t=t_{\mathrm{out}}$ reads%
\begin{equation*}
\max \mathcal{E}_{2}(t_{\mathrm{out}})\approx \left\{ 
\begin{array}{c}
\frac{1}{2}\left\vert g(t_{\mathrm{out}})\right\vert ,\ \ \mathrm{if\ }%
\sigma =-1, \\ 
\frac{3}{4}\left\vert g(t_{\mathrm{out}})\right\vert ,\ \ \mathrm{if\ }%
\sigma =+1.%
\end{array}%
\right.
\end{equation*}%
However, during the stage of $\nu _{e}\bar{\nu}_{e}$ pair creation with $%
\sigma =+1$, $t<t_{0}$, when $g_{2}=g(t)>0$, the maximal kinetic energy of
created electron neutrino or antineutrino increases as $\max \mathcal{E}%
_{2}\left( t\right) \approx \frac{1}{2}\left[ g(t_{\mathrm{in}})-g(t)\right] 
$ and reaches its maximal value $\max \mathcal{E}_{2}\left( t_{0}\right)
\approx \frac{1}{2}g(t_{\mathrm{in}})=\frac{1}{4}\left\vert g(t_{\mathrm{out}%
})\right\vert $ at the end of this stage. This value of the maximal kinetic
energy is consistent with the fact that the rest $\frac{1}{2}\left\vert g(t_{%
\mathrm{out}})\right\vert $ of the final kinetic energy $\max \mathcal{E}%
_{2}(t_{\mathrm{out}})$ of this neutrino is gained due to the acceleration
of already existing particle after the time instant $t_{0}$.

Thus, total numbers of states $\Delta _{\sigma }^{e}$ of the electron
neutrino with a fixed helicity in the momentum range given by Eqs.~(\ref%
{4.18}) or (\ref{4.19}) are 
\begin{align}
\Delta _{-1}^{e}=&\frac{V\left\vert g(t_{\mathrm{out}})\right\vert ^{3}}{%
3\left( 4\pi \right) ^{2}}\left[ 1+O\left( \frac{\sqrt{a}K}{\left\vert g(t_{%
\mathrm{out}})\right\vert }\right) \right] ,
\notag
\\
\Delta _{+1}^{e}=&\frac{V%
\left[ g(t_{\mathrm{in}})\right] ^{3}}{3\left( 4\pi \right) ^{2}}\left[
1+O\left( \frac{\sqrt{a}K}{g(t_{\mathrm{in}})}\right) \right] .  \label{4.20}
\end{align}%
We see that $\Delta _{-1}^{e}=8\Delta _{+1}^{e}$. Using Eq.~(\ref{4.17.2})
and (\ref{4.20}), we find the energy density of created neutrinos or
antineutrinos with a given helicity,%
\begin{align}
w_{-1}^{e} = &\left\langle \mathcal{E}_{-1}^{e}\right\rangle e^{-\pi \lambda
_{e}}\Delta _{-1}^{e}/V,
\notag
\\
\left\langle \mathcal{E}_{-1}^{e}\right\rangle =& %
\frac{1}{8}\left\vert g(t_{\mathrm{out}})\right\vert ,\;\;\mathrm{if}%
\;\;\sigma =-1,  \notag \\
w_{+1}^{e} =&\left\langle \mathcal{E}_{+1}^{e}\right\rangle e^{-\pi \lambda
_{e}}\Delta _{+1}^{e}/V,
\notag
\\
\left\langle \mathcal{E}_{+1}^{e}\right\rangle =&%
\frac{11}{16}\left\vert g(t_{\mathrm{out}})\right\vert ,\;\;\mathrm{if}%
\;\;\sigma =+1,  \label{4.21}
\end{align}%
where $\left\langle \mathcal{E}_{\sigma }^{e}\right\rangle $ is the mean
energy per an electron neutrino or an antineutrino created. We see that the
mean energy $\left\langle \mathcal{E}_{-1}^{e}\right\rangle $ is much less
than $\left\langle \mathcal{E}_{+1}^{e}\right\rangle $, though the energy
densities of created electron neutrinos with the opposite helicity are of
the same order, $w_{+1}^{e}=$ $\frac{11}{16}w_{-1}^{e}$.

For $\nu _{\mu ,\tau }$ it follows from Eq.~(\ref{1.4}) that $g_{1}=g(t_{%
\mathrm{in}})<0$ and $g_{2}=g(t_{\mathrm{out}})=+2g(t_{\mathrm{in}})$. Using
Eq.~(\ref{4.15}), we find that in the momentum range%
\begin{equation}
D_{-1}^{\mu ,\tau }:\frac{\left\vert g_{1}\right\vert }{2}+\sqrt{\frac{a}{2}}%
K\leq p\leq \frac{\left\vert g_{2}\right\vert }{2}-\sqrt{\frac{a}{2}}K
\label{4.22}
\end{equation}%
the only $\nu _{\mu ,\tau }\bar{\nu}_{\mu ,\tau }$ pairs with $\sigma =-1$
are created. The maximal kinetic energy of $\nu _{\mu ,\tau }$ or $\bar{\nu}%
_{\mu ,\tau }$ neutrinos created at final time instant $t=t_{\mathrm{out}}$
follows from Eq.~(\ref{4.22}) to be $\max \mathcal{E}_{2}(t_{\mathrm{out}%
})\approx \frac{1}{4}\left\vert g(t_{\mathrm{out}})\right\vert $. In the
same range, the total number of $\nu _{\mu ,\tau }$ neutrino states with $%
\sigma =-1$ has the form%
\begin{align}
\Delta _{-1}^{\mu ,\tau }=&\left\{ \frac{V\left\{ \left\vert g(t_{\mathrm{out}%
})\right\vert ^{3}-\left\vert g(t_{\mathrm{in}})\right\vert ^{3}\right\} }{%
3\left( 4\pi \right) ^{2}}\right\} 
\notag
\\
&\times
\left[ 1+O\left( \frac{\sqrt{a}K}{g(t_{%
\mathrm{in}})}\right) \right] .  \label{4.23}
\end{align}%
The energy density and the mean energy per a particle for created $\nu _{\mu
,\tau }\bar{\nu}_{\mu ,\tau }$ are, respectively,%
\begin{align}
w_{-1}^{\mu ,\tau }=&\left\langle \mathcal{E}_{-1}^{\mu ,\tau }\right\rangle
e^{-\pi \lambda _{\mu ,\tau }}\Delta _{-1}^{\mu ,\tau }/V,
\notag
\\
\left\langle 
\mathcal{E}_{-1}^{\mu ,\tau }\right\rangle =&\frac{11}{112}\left\vert g(t_{%
\mathrm{out}})\right\vert .  \label{4.24}
\end{align}

The effective potential for $\nu _{e}$\ does not change at the compression
stage then there is no $\nu _{e}\bar{\nu}_{e}$\ creation. Just as the
initial $g_{1}=g(t_{\mathrm{in}})\approx 0$\ and the final $g_{2}=g(t_{%
\mathrm{out}})<0$\ for $\nu _{\mu ,\tau }$\ at this stage. Using Eq.~(\ref%
{4.15}), we find that the only $\nu _{\mu ,\tau }\bar{\nu}_{\mu ,\tau }$\
pairs with $\sigma =-1$\ are created due to the compression and the range of
momenta of these pairs is%
\begin{equation}
D_{-1}^{\mu ,\tau }\left( \mathrm{c}\right) :\sqrt{\frac{a}{2}}K\leq p\leq 
\frac{\left\vert g_{2}\right\vert }{2}-\sqrt{\frac{a}{2}}K.  \label{4.25}
\end{equation}%
The maximal kinetic energy of created particles at final time instant is $%
\max \mathcal{E}_{2}(t_{\mathrm{out}})\approx $\ $\frac{1}{2}\left\vert g(t_{%
\mathrm{out}})\right\vert $\ and the total number of states that belong to
the range (\ref{4.25}) is%
\begin{equation}
\Delta _{-1}^{\mu ,\tau }\left( \mathrm{c}\right) =\frac{V\left\vert g(t_{%
\mathrm{out}})\right\vert ^{3}}{3\left( 4\pi \right) ^{2}}\left[ 1+O\left( 
\frac{\sqrt{a}K}{\left\vert g(t_{\mathrm{out}})\right\vert }\right) \right] .
\label{4.26}
\end{equation}%
Then the energy density and the mean energy per a particle for created $\nu
_{\mu ,\tau }\bar{\nu}_{\mu ,\tau }$\ at final time of the compression are,
respectively,%
\begin{align}
w_{-1}^{\mu ,\tau }\left( \mathrm{c}\right) =&\left\langle \mathcal{E}%
_{-1}^{\mu ,\tau }\left( \mathrm{c}\right) \right\rangle e^{-\pi \lambda
_{\mu ,\tau }}\Delta _{-1}^{\mu ,\tau }\left( \mathrm{c}\right)
/V,
\notag
\\
\left\langle \mathcal{E}_{-1}^{\mu ,\tau }\left( \mathrm{c}\right)
\right\rangle =&\frac{1}{8}\left\vert g(t_{\mathrm{out}})\right\vert .
\label{4.27}
\end{align}

Assuming that $g_{2}=g\left( t\right) $ changes from $g(t_{\mathrm{in}})$\ \
to $g(t_{\mathrm{out}})$, one can obtain time dependence of all the physical
quantities during the neutronization. Note that the numbers of states $%
\Delta _{\sigma }$ given by Eqs.~(\ref{4.20}), (\ref{4.23}), and (\ref{4.26}%
) are nonlinear functions of the time instants $t_{\mathrm{out}}$ and $t_{%
\mathrm{in}}$. Therefore the total particle production rate is not a
conserved physical quantity in this case.

\section{Neutrino creation in realistic astrophysical media\label{S5}}

In this section, in the framework of the above developed technique we study $%
\nu \bar{\nu}$ pair creation in realistic astrophysical media. In
particular, we consider this effect\textbf{\ }at the compression stage
before the hydrodynamic bounce and at the neutronization of PNS. In both
cases we derive the upper limit on neutrino masses that corresponds to the
nonvanishing probability of $\nu \bar{\nu}$ pairs creation. Then we discuss
the evolution of the created neutrinos.

It is commonly believed that a star having ($10-25$) solar masses, ends its
evolution as a neutron star through a core-collapsing supernova stage with
the emission of $99\%$ of the initial gravitational energy in the form of
neutrinos~\cite{Heg03}.

According to the modern simulations (see, e.g., Ref.~\cite{Jan07}) the
density in the central part on PNS reaches $\sim 10^{12}\,\text{g}\cdot 
\text{cm}^{-3}$ at $\sim 100\,\text{s}$ after the beginning of the collapse.
High-energy ($E \geq 10\thinspace\text{MeV}$) neutrinos, which are created
in the core of PNS, cannot escape since their mean free path is much less
than the core radius. During the next $T_{\nu }\approx 10\,\text{ms}$ the
central density increases to $\gtrsim 2\times 10^{14}\,\text{g}\cdot \text{cm%
}^{-3}$. At this stage the compression of matter in PNS core stops and the
hydrodynamic bounce happens.

The bounce is typically followed by the neutronization of PNS matter. The
neutronization is characterized by the change of $Y_{e}$ from $0.5$ to
practically zero value. This process occurs outside the PNS core at $10\,%
\text{km}\lesssim r\lesssim 100\,\text{km}$, begins at $t\approx 0.11\,\text{%
s}$, and lasts $T_{\nu }\sim 10^{-2}\,\text{s}$ (see, e.g., Ref.~\cite{Jan07}%
). The liberated lepton number is carried away by $\nu _{e}$ produced in the
reaction $e^{-}+p\rightarrow n+\nu _{e}$ and having the energy $\sim 10\,%
\text{MeV}$.

First let us we discuss the creation of $\nu \bar{\nu}$ pairs \ due to the
matter compression using our formalism during $T_{\nu }=10\,\text{ms}$ just
before the bounce. We should mention that one can neglect the radial
hydrodynamic currents directed towards the center of PNS [see Eq.~%
\eqref{efPot}] in the effective potential of the neutrino interaction with
background fermions. Such a contribution is inevitable since the central
density is increasing. Supposing that all background fermions have
approximately equal radial velocities $v_{r}$ and using Eq.~\eqref{efPot} we
get that $g_{r}/g=v_{r}$. As found in Ref.~\cite{ThoBurPin03}, $%
v_{r}\lesssim 0.1$ inside the PNS core, $r\lesssim 10\,\text{km}$, within
the considered time of the PNS evolution. Thus $g_{r}$ is much less than $g$.

Since the matter density increases two orders of magnitude, we can take that 
$g(t_\mathrm{in}) \approx 0$. The electron fraction $Y_e = n_e/(n_n + n_p)$
changes from $\sim 0.5$ to $\sim 0.3$~\cite{ThoBurPin03}, which corresponds
to $n_n(t_\mathrm{out}) \approx 2 n_e(t_\mathrm{out})$. Therefore, using Eq.~%
\eqref{eq:gexpl} we get that $g(t_\mathrm{out}) \approx 0$ for $\nu_e$. Thus
the creation of $\nu_e \bar{\nu}_e$ pairs is suppressed at this stage of the
PNS evolution.

Again using Eq.~\eqref{eq:gexpl} we get that for $\nu_{\mu,\tau}$, 
\begin{align}
g_{1}=&g(t_{\mathrm{in}}) = 0,
\notag
\\
g_{2}=&g(t_{\mathrm{out}})=-G_{\mathrm{F}%
}n_{n} \left( t_{\mathrm{out}} \right) /\sqrt{2}.
\end{align}
Therefore $\Delta g_{\nu_{\mu,\tau}} = |g_1 - g_2| \neq 0$ and the creation
of $\nu_{\mu,\tau} \bar{\nu}_{\mu,\tau}$ pairs is possible. We shall roughly
assume that the effective potential changes linearly from zero to $g_2$.
Thus the results of Sec.~\ref{S4} are valid.

It results from Eqs.~(\ref{4.15a}) and (\ref{4.16}) that the flux of
low-energy $\nu _{\mu ,\tau }\bar{\nu}_{\mu ,\tau }$ pairs is sizable if $%
\lambda _{\nu _{\mu ,\tau }}=2m_{\nu _{\mu ,\tau }}^{2}T_{\nu }/\Delta
g_{\nu _{\mu ,\tau }}\lesssim 1$. Assuming that $Y_{e}\approx 1/3$, $\rho
=2\times 10^{14}\,\text{g}\cdot \text{cm}^{-3}$, and $T_{\nu }=10^{-2}\,%
\text{s}$, we get that $\Delta g_{\nu _{\mu ,\tau }}\approx 5\,\text{eV}$,
where we use value of the Fermi constant $G_{\mathrm{F}}(\hbar
c)^{-3}\approx 1.17\times 10^{-5}\,\text{GeV}^{-2}$. Finally we obtain the
constraint on the electron neutrino mass, 
\begin{equation}
m_{\nu _{\mu ,\tau }}\lesssim m_{\nu _{\mu ,\tau }}^{\mathrm{({cr})}%
}=4.1\times 10^{-7}\,\text{eV}.  \label{mcrbounce}
\end{equation}%
Note that, if the constraint in Eq.~\eqref{mcrbounce} is fulfilled, the flux
of $\nu _{\mu ,\tau }\bar{\nu}_{\mu ,\tau }$ pairs is nonvanishing.

It is interesting to mention that, in the considered time interval just
before the bounce, high-energy neutrinos are produced in the PNS core.
However these neutrinos are trapped inside the core due to elastic and
quasielastic neutrino scattering off background fermions. We predict a
nonzero flux of $\nu _{\mu ,\tau }\bar{\nu}_{\mu ,\tau }$ pairs having very
small energy $<10\,\text{eV}$. These neutrinos are not trapped inside the
core. % due to the above mentioned scattering off background fermions.
Indeed, using the neutrino scattering cross sections given in Ref.~\cite%
{GiuKim07p160}, one finds that the mean free path of these particles in
background matter with the density $2\times 10^{14}\,\text{g}\cdot \text{cm}%
^{-3}$ is about $10^{10}\,\text{km}$. Therefore one can consider these
neutrinos as precursors of neutronization neutrino burst.

Now let us consider the $\nu \bar{\nu}$ pairs creation during the
neutronization of PNS. Since the number densities of various background
fermions change, with the total mass density of PNS matter being constant,
the effective potential in Eq.~(\ref{eq:gexpl}) also changes [see, e.g., Eq.~%
\eqref{1.4}] and we may expect that an additional flux of low-energy $\nu 
\bar{\nu}$ pairs can be emitted at the neutronization of PNS. Again we shall
assume that the effective potential changes linearly.

(i) First we suppose that the electron neutrino mass $m_{\nu _{e}}$ is the
smallest among the all neutrino masses.

Since $Y_{e}$ changes from $0.5$ to $0$ in the neutronization of PNS, the
number densities before and after the neutronization satisfy, $n_{e}\left(
t_{\mathrm{in}}\right) \approx n_{n}\left( t_{\mathrm{in}}\right) $ and $%
n_{n}\left( t_{\mathrm{out}}\right) =n_{n}\left( t_{\mathrm{in}}\right)
+n_{e}\left( t_{\mathrm{in}}\right) \approx 2n_{n}\left( t_{\mathrm{in}%
}\right) $ Therefore, using Eq.~(\ref{eq:gexpl}), we obtain 
\begin{align}
g_{1}=&g(t_{\mathrm{in}})=G_{\mathrm{F}}n_{n}\left( t_{\mathrm{out}}\right)
/(2\sqrt{2}),
\notag
\\
g_{2}=&g(t_{\mathrm{out}})=-G_{\mathrm{F}}n_{n}\left( t_{%
\mathrm{out}}\right) /\sqrt{2},
\end{align}%
such that for the electron neutrino we have $\Delta g_{\nu _{e}}=\left\vert
g_{2}-g_{1}\right\vert =\frac{3}{2}\left\vert g(t_{\mathrm{out}})\right\vert 
$ .

Requiring the nonvanishing flux of $\nu _{e}\bar{\nu}_{e}$ pairs by imposing 
$\lambda _{\nu _{e}}=2m_{\nu _{e}}^{2}T_{\nu }/\Delta g_{e}\lesssim 1$ [see
Eqs.~(\ref{4.15a}) and (\ref{4.16})], we get the constraint on the electron
neutrino mass,%
\begin{equation}
m_{\nu _{e}}\lesssim m_{\nu _{e}}^{\mathrm{({cr})}}=5.6\times 10^{-8}\,\text{%
eV}.  \label{5.1}
\end{equation}%
To derive Eq.~\eqref{5.1} we assume that $T_{\nu }=10^{-2}\,\text{s}$ and $%
n_{n}(t_{\mathrm{out}})\sim 10^{36}\,\text{cm}^{-3}$ then\textbf{\ }$g(t_{%
\mathrm{out}})\approx 0.064~\mathrm{eV}$\textbf{. }The latter quantity
corresponds to the mass density $\lesssim 10^{12}\,\text{g}\cdot \text{cm}%
^{-3}$.

(ii) Now we suppose that the smallest among the all neutrino masses is
either $m_{\nu _{\mu}} $ or $m_{\nu _{\tau }}$.

The treatment of both muon and tau neutrinos is the same. For $\nu _{\mu }$
and $\nu _{\tau }$ we get from Eq.~(\ref{1.4}) that unlike the case (i) the
initial and final effective potentials are 
\begin{align}
g_{1}=&g(t_{\mathrm{in}})=-G_{\mathrm{F}}n_{n}\left( t_{\mathrm{out}}\right)
/(2\sqrt{2}),
\notag
\\
g_{2}=&g(t_{\mathrm{out}})=+2g(t_{\mathrm{in}}).
\end{align}%
Therefore $\Delta g_{\nu _{\mu ,\tau }}=\left\vert g_{2}-g_{1}\right\vert =%
\frac{1}{2}\left\vert g(t_{\mathrm{out}})\right\vert $. The flux of the
low-energy $\nu _{\mu ,\tau }\bar{\nu}_{\mu ,\tau }$ pairs is big enough if $%
\lambda _{\mu ,\tau }=2m_{\nu _{\mu ,\tau }}^{2}T_{\nu }/\Delta g_{\mu ,\tau
}\lesssim 1$. Thus, we obtain the constraint on the appropriate muon and tau
neutrino masses:%
\begin{equation}
m_{\nu _{\mu ,\tau }}\lesssim m_{\nu _{\mu ,\tau }}^{\mathrm{({cr})}%
}=1.9\times 10^{-8}\,\text{eV}.  \label{5.2}
\end{equation}%
It should be also noted that the energy of these $\nu \bar{\nu}$\ pairs does
not exceed $0.1~\text{eV }$\ for all three active neutrino flavors. The
dimensionless parameters in Eq.~(\ref{4.11}) are quite large for the cases
(i) and (ii) at the neutronization, $\left( \left\vert
g_{2}-g_{1}\right\vert T_{\nu }\right) ^{1/2}\sim 10^{6}$, and for the
compression,\textbf{\ }$\sim 10^{7}$. Then this condition is well satisfied
for the subcritical masses given by Eqs.~(\ref{mcrbounce}), (\ref{5.1}),
and~(\ref{5.2}).

In Appendix~\ref{SS5.2} we analyzed the influence of other factors which can
diminish the flux of created $\nu \bar{\nu}$ pairs or distort their
distribution. Among them we considered the possible Pauli blocking of the
creation process, the gravitational interaction of the low-energy neutrinos,
the influence of the PNS rotation on the pairs propagation, and low-energy
pair production by nucleon-nucleon bremsstrahlung. We found that all these
processes do not significantly influence the evolution of $\nu \bar{\nu}$
pairs created \ if the mass of the neutrino is small enough. The only factor
which essentially influences the evolution of $\nu \bar{\nu}$ pairs created
is the difference between the effective density in the region of the
creation and in the point outside this region. The high-density region is a
potential well for either neutrino or antineutrino depending on the sign of
the effective potential. Then part of these particles, depending on the
flavor and helicity, are bounded in the PNS while the antineutrinos of any
flavor escape the PNS. If the created pairs are $\nu _{e}\bar{\nu}_{e}$\
then part of these neutrinos also escape the PNS. A part of escaped
neutrinos that have the negative helicity can interact directly with the
matter of electrons and baryons. All the escaped antineutrinos have the
negative helicity and do not interact directly with the uniform part of the
matter consisting of electrons and baryons. Nevertheless, an effective
potential barrier of a neutron star can affect them, causing refraction and
reflection, and, in particular, change their helicity in course of a
reflection.

Additionally, we evaluated the typical flux of neutrino/antineutrino,
created in frames of our formalism, from a possible supernova in our Galaxy,
which can reach the Earth. We considered pairs emitted during the core
compression stage which have $g(t_{\mathrm{out}})\sim 10\,\text{eV}$ and the
numbers of occupied states $\Delta _{-1}^{\mu ,\tau }\left( \mathrm{c}%
\right) $ given by Eq.~(\ref{4.26}) is of the order of $\frac{V\left\vert
g(t_{\mathrm{out}})\right\vert ^{3}}{3\left( 4\pi \right) ^{2}}\sim 10^{33}$
for $R_{\mathrm{c}}\sim 10~\text{km}$. Supposing that the distance to a
supernova $\sim 1\,\text{kpc}$ and a potential detector has the effective
area $1\,\text{km}^{2}$, we get that about $10$ particles could interact
with such a detector. In this case the counting rate is $\sim 10^{6}\,\text{s%
}^{-1}$. And the typical flux created at the neutronization stage is $10^{3}$
times smaller. The obtained quantity is much smaller than an expected
counting rate of high-energy neutrinos from our Galaxy supernova.

The estimates of the neutrino masses given in Eqs.~\eqref{mcrbounce}, (\ref%
{5.1}), and (\ref{5.2}) does not contradict the modern constraints on the
neutrino masses (see, e.g., Ref.~\cite{Ase11}). Of course, direct detecting
such low-energy neutrinos or antineutrinos is beyond any existing
experimental possibilities. The total energy radiated of these neutrino is
about $10^{22}~\mathrm{erg}$. This is a completely negligible amount of
energy compared to other scales in the supernova problem or in relation to
the energy scales in the outer layers of the star. Hence, this flux of
created $\nu \bar{\nu}$\ pairs cannot affect the evolution of the star and
shows its presence by such a way.\ Since the flux of low-energy $\nu \bar{\nu%
}$ pairs from a supernova has not been detected yet, our constraints on
neutrino masses should be regarded as a condition for the creation of a
nonvanishing flux of neutrino pairs in matter with the time-dependent
effective potential. \ 

Note that the $\nu \bar{\nu}$ pair creation from the vacuum considered in
the present work is the result of a unitary evolution. As a consequence,
low-energy particles are coherently emitted in a macroscopic region. The
flux of low-energy neutrinos predicted in our work will be accompanied by
the radiation of high-energy neutrinos. However, the spectra of highly
energetic $\nu _{e,\mu ,\tau }$\ and $\bar{\nu}_{e,\mu ,\tau }$\ emitted at
the neutronization stage of PNS are pinched at low and high-energy parts
relative to the mean energy $\sim 10\,\text{MeV}$ (see details in Appendix~%
\ref{SS5.2}). That is, the very rare $\nu $\ and $\bar{\nu}$\ of such origin
can lose enough part of their energy during neutronization to get the
considered low-energy range and these particles, produced independently in
the reaction between several particles, are statistically independent. \ It
means that, in principle,\textbf{\ }particles emitted coherently are
statistically distinguishable from the latter.\textbf{\ }The length scale,
associated with $T_{\nu }$ is $\sim 10^{8}\,\text{cm}$, which is much bigger
than both the PNS core radius $R_{\mathrm{c}}\sim 10\,\text{km}$ and the
radius of the sphere where the neutronization happens $R_{\mathrm{n}}\sim
100\,\text{km}$. Thus, PNS will be a coherent source of low-energy $\nu \bar{%
\nu}$ pairs. These low-energy neutrinos may be involved in some interference
effects, e.g., in their lensing by the effective potential barriers of
neutron stars and gravity. If we hypothesize that the detection of
low-energy neutrinos is possible due to yet unknown mechanism for resonance
amplification of the signal, these effects can help one to separate such
coherent fluxes from chaotic fluxes of other origin. Currently detecting
such low-energy $\nu $\ and $\bar{\nu}$\ seems to be impossible despite the
recent theoretical proposals of corresponding experiments of the $\text{meV}$%
\ energy scale, see, for example Ref.~\cite{Yosh07,KATRIN13}.

\section{Summary\label{S6}}

In this summary we briefly list the main new results obtained in the present
work and organize them conditionally into the following three blocks:

(i) We have considered the Dirac neutrino interacting with background
fermions in the frame of the standard model. We demonstrate that a
time-dependent effective potential is quite possible in a protoneutron star
(PNS) at the compression stage just before the hydrodynamic bounce and
during PNS neutronization. Such an interaction is intense and must be
treated nonperturbatively.

For the first time, we have formulated in the framework of the quantum field
theory a corresponding nonperturbative treatment of neutrino processes in a
matter with arbitrary time-dependent effective potential. This allowed us to
study analytically a realistic case of slowly varying effective potential.
Using complete sets of exact solutions of the Dirac equation in the
time-dependent effective potential, we have constructed the initial and
final Fock spaces and Bogolyubov transformations between the corresponding
creation and annihilation operators. We have expressed mean numbers of $\nu 
\bar{\nu}$ pairs created from the vacuum and the probabilities of all the
transition processes via coefficients in the Bogolyubov transformations.

(ii) A model with linearly and slowly growing effective potential that has a
large difference of its initial and final values compared with the neutrino
mass was studied in detail. It was shown that results obtained for this
model are representative for a large class of slowly varying potentials.%
\emph{\ }We have calculated differential mean numbers of $\nu \bar{\nu}$
pair created from the vacuum and have found that they crucially depend on
the effective mass of a lightest neutrino. These distributions uniformly
span from $\sim 10^{-6}~\mathrm{eV}\,$ to $\sim 10~\mathrm{eV}$ energies for 
$\nu _{\mu ,\tau }\bar{\nu}_{\mu ,\tau }$ created due to the compression and
from $\sim 10^{-6}~\mathrm{eV}$ to $\sim 0.1~\mathrm{eV}$ energies for all
three active neutrino flavors created due to the neutronization dropping
sharply beyond this interval. We have obtained the total number and the
energy density of created $\nu \bar{\nu}$ pairs and examined peculiarities
in the production of different neutrino flavors and helicities.

(iii) We have studied $\nu \bar{\nu}$ pair production from vacuum in a PNS
core at the compression stage just before the hydrodynamic bounce and during
the PNS neutronization. It was shown that the creation of pairs of
low-energy neutrinos up to $\sim 10~\mathrm{eV}$ is possible in these cases.
These low-energy pairs are coherently emitted from a macroscopic region
during the considered stages of the PNS evolution. Part of these particles,
depending on the flavor and helicity, are bounded in the PNS while the
antineutrinos of any flavor escape the PNS. If the created pairs are $\nu
_{e}\bar{\nu}_{e}$\ then part of these neutrinos also escape the PNS. Only a
part of these escaped neutrinos interacts directly with the uniform matter
of electrons and baryons. In general, an effective potential barrier of a
neutron star can affect such low-energy neutrinos and antineutrinos, causing
refraction and reflection, and, in particular, change their helicity in
course of a reflection. Thus, accounting for the characteristic isotropic
uniform distribution of $\nu \bar{\nu}$ pairs created in the low-energy
range and specific properties dependent on the neutrino flavors, we have
shown that one can distinguish such coherent flux from chaotic fluxes of any
other origin.\emph{\ }We have derived constraints on the neutrino masses: $%
m_{\nu _{\mu ,\tau }}\lesssim 4.1\times 10^{-7}\mathrm{eV}$, for particles
created in the core compression before the bounce, as well as $m_{\nu
_{e}}\lesssim 5.6\times 10^{-8}\,\mathrm{eV}$ and $m_{\nu _{\mu ,\tau
}}\lesssim 1.9\times 10^{-8}\,\mathrm{eV}$ for the pairs emission at the
neutronization, corresponding to the nonvanishing $\nu \bar{\nu}$ pairs flux
produced by this mechanism. We have examined other processes which might
affect detection of this vacuum instability in the PNS and found that they
are negligible if the mass of the neutrino is small enough. The energies of
created neutrinos are less than $10\,\text{eV}$, for particles emitted
before the bounce, and less than $0.1\,\text{eV}$, for the emission at the
PNS neutronization. We should mention that $\bar{\nu}_{\mu ,\tau }$ of the
pairs created before the bounce freely escape the dense core unlike their
high-energy counterparts. Thus these particles can be regarded as precursors
of the neutronization neutrino burst. Unfortunately, current experimental
techniques do not allow one to detect neutrinos with such low energies.

\begin{acknowledgments}
M.D. is indebted to FAPESP (Brazil) for a grant and to Y.~Kivshar for the
hospitality at the ANU where a part of the work was made. S.P. Gavrilov
thanks FAPESP for a support and University of S\~{a}o Paulo for the
hospitality. D. Gitman thanks CNPq and FAPESP for permanent support.
\end{acknowledgments}

\appendix

\section{ACCOMPANYING PROCESSES\label{SS5.2}}

In this Appendix we consider possible processes which might affect either
the creation of the neutrino pairs or their subsequent propagation at the
initial stages of the PNS evolution. The creation of $\nu _{\mu ,\tau }\bar{%
\nu}_{\mu ,\tau }$\ pairs \ due to the matter compression and their
propagation occur before the neutronization. Thus the accompanying processes
which can infuence these two phenomena do not overlap.

Concerning $\nu \bar{\nu}$ pairs created at the neutronization, we can
conclude the following. We obtain from Eq.~(\ref{3.21}) that a filled
neutrino and/or antineutrino initial state blocks the neutrino creation with
the corresponding quantum number. However, we see no reason to expect that
the occupation numbers of the initial distribution $\aleph _{n}^{(\zeta )}(%
\mathrm{in})$ in the range of low energies being uniformly great immediately
after the start of a neutronization stage. As found in Ref.~\cite{Tot98},
the spectra of highly energetic $\nu _{e,\mu ,\tau }$ and corresponding
antiparticles emitted at the neutronization stage of PNS are not Fermi-Dirac
ones. In particular these spectra are pinched at low- and high- energy parts
relative to the mean energy $\sim 10\,\text{MeV}$. For $\nu _{e}$ and $\bar{%
\nu}_{e}$ the relaxation time to reach the thermal distribution is longer
than $T_{\nu }$~\cite{ThoBurPin03,Tot98}. It was revealed in Ref.~\cite%
{ThoBurHor00} that for other neutrino species the relaxation time also
exceeds $T_{\nu }$. Therefore we get that the creation of low-energy $\nu 
\bar{\nu}$ pairs by our mechanism cannot be suppressed by the Pauli factor
since the lowest energy states are unoccupied.

It should be noted that besides the $\nu \bar{\nu}$ pair creation by the
spatially homogeneous effective potential $g(t)$ at the neutronization
stage, we can expect that the inhomogeneity of the PNS matter will affect
the propagation of low-energy neutrinos escaping the PNS. Let us examine
this effect.

For the case of the matter compression,\textbf{\ }we may roughly assume that
the core of PNS has an approximately constant density with $n_{n}\sim
10^{38}\,\text{cm}^{-3}$. The PNS core density decreases several orders of
magnitude in the spherical PNS crust which has the thickness $\Delta R_{%
\mathrm{c}}\sim 1\,\text{km}$~\cite{HaePotYak07p11}. Taking into account the
range of neutrino momenta under consideration given in Eq.~(\ref{4.25}), we
see that all low-energy neutrinos and antineutrinos are ultrarelativistic
particles. It takes $\sim 10^{-6}\,\text{s}$ for such particles to pass
through the PNS crust. The PNS density of the spherical shell of the
neutronization,\textbf{\ }$10\,\text{km}\lesssim r\lesssim 100\,\text{km}$%
\textbf{, }is of the order of\textbf{\ }$n_{n}\sim 10^{36}\,\text{cm}^{-3}$.%
\textbf{\ }One can assume that the density of this shell decreases
significantly at a distance of $\sim 10\,$km near the outer boundary, $%
\Delta R_{\mathrm{n}}\sim 10\,$km. The neutrino and antineutrino created due
to neutronization are ultrarelativistic particles as well. It takes $\sim
10^{-5}\,$s for such particles to escape through the $10\,$km\ thickness of
the outer shell of significant gradient. Both time scales are much shorter
than $T_{\nu }$. Therefore we can consider process of the inhomogeneity
region crossing as independent one.

To analyze this process we can assume that the effective matter density $g_{%
\mathrm{int}}=g\left( t\right) $ in the shells of significant gradient
varies adiabatically from $g(t_{\mathrm{in}})$\ \ to $g(t_{\mathrm{out}})$
and the corresponding gradient of the effective matter density varies
smoothly. It is worth mentioning that the size of the wave packet of the
low-energy neutrinos under consideration is in the range $\sim
(10^{-5}-10^{2})~\text{cm,}$ which is much smaller than the scale of the
matter inhomogeneity.

One can accordingly describe the macroscopic part of these shells using the
time-independent one dimensional effective matter density $g\left( r\right) $
that depends only on a radial coordinate $r$ orthogonal to the border and
represents a kind of potential step. We assume that the density $g\left(
r\right) $ varies smoothly from the value $g_{\mathrm{int}}$ in the core to $%
g_{\mathrm{ext}}=0$ in the space outside the shell under consideration with
a constant gradient $a^{\prime }=-g_{\mathrm{int}}/\Delta R$.

Thus one can treat the effect of the border using the Dirac equation (\ref%
{eq:Direq}) with the matter density $g\left( X\right) =g\left( r\right) $.
Such an equation is quite similar to the Dirac equation for the electron in
an electric field given by scalar step potential, where $g\left( r\right) /2$
and $a^{\prime }/2$ play roles of these potential and constant electric
field, respectively. The gradient $\left\vert a^{\prime }\right\vert $ is
considerably larger than above mentioned $\left\vert a\right\vert \sim
\left( m_{\nu _{e,\mu ,\tau }}^{\mathrm{({cr})}}\right) ^{2}$ during the%
\textbf{\ }compression stage, $\left\vert a^{\prime }\right\vert \sim
10^{4}\left\vert a\right\vert $, and during the neutronization, $\left\vert
a^{\prime }\right\vert \sim 10^{2}\left\vert a\right\vert $. Hence such a
field is very strong for the both subcritical masses given by Eq.~(\ref%
{mcrbounce}), $\left( m_{\nu _{\mu ,\tau }}^{\mathrm{({cr})}}\right) ^{2}\ll
\left\vert a^{\prime }\right\vert $, and Eqs.~(\ref{5.1}) and (\ref{5.2}), $%
\left( m_{\nu _{e,\mu ,\tau }}^{\left( \mathrm{cr}\right) }\right) ^{2}\ll
\left\vert a^{\prime }\right\vert $, respectively.

The similar problem of the $\nu \bar{\nu}$ pairs creation from vacuum in
cold neutron stars was considered in Refs.~\cite{Loeb90,Kach98} and the
production rate of $\nu \bar{\nu}$ pairs is evaluated following an analogy
with Schwinger's result for $e^{+}e^{-}$ creation by a constant uniform
electric field \cite{schwinger}. This approach is not applicable for our
problem since it does not allows us to estimate the mean number of particles
created within a finite time $T$ on a finite length $\Delta R$. In our case
a more detailed analysis is required, analogous to that made in Refs.~\cite%
{Nik-barrier-70,Nikis79,WongW88} where the $e^{+}e^{-}$ pair creation by a
constant uniform electric field given by scalar potential was studied.

The differential mean number of neutrino or antineutrino created from vacuum
by the inhomogeneous matter can be evaluated in analogy with the case of the
electric field, yielding%
\begin{equation}
N_{n}^\mathrm{gr}\approx \exp \left[ -2\pi \left( m^{2}+\mathbf{p}_{\bot
}^{2}\right) /\left\vert a^{\prime }\right\vert \right] ,  \label{5.3}
\end{equation}%
where $m$ is the corresponding neutrino mass, $n=\left( p_{0},\mathbf{p}%
_{\bot },s\right) $ is the complete set of quantum numbers, $p_{0}$ is the
total energy, $\mathbf{p}_{\bot }$ is transversal momentum that is
orthogonal to the gradient direction, and $s$ is a given spin polarization.
Note that the distribution $N_{n}^\mathrm{gr}$ decreases very rapidly with
increasing transversal momentum.

It can be shown that the expression given by Eq.~(\ref{5.3}) is valid in the
range of the energy $\left\vert p_{0}\right\vert <$ $\left\vert g_{\mathrm{%
int}}\right\vert /2$ and the value of $N_{n}^{\mathrm{gr}}$ is negligible
outside this range. The accurate nonperturbative treatment of $\nu \bar{\nu}$
pairs creation due to the inhomogeneity of the matter density can be
performed using the formalism recently developed in Ref.~\cite{GavGit13a}.
The appropriate general QFT formalism is developed in Ref.~\cite{GavGit13b}.
Note that the value given by Eq.~(\ref{5.3}) saturates for low values of $%
\mathbf{p}_{\bot }^{2}$, $N_{n}^{\mathrm{gr}}\simeq 1$ for the subcritical
masses, $m\lesssim m_{\nu _{e,\mu ,\tau }}^{\left( \mathrm{cr}\right) }$.The
total number of particles created by this mechanism can be found as%
\begin{equation}
\mathcal{N}^{\mathrm{gr}}\approx \frac{T_{\nu }S_{R}}{\left( 2\pi \right)
^{3}}\sum_{s=\pm 1}\int N_{n}^{\mathrm{gr}}dp_{0}d\mathbf{p}_{\bot },
\label{5.4}
\end{equation}%
where $S_{R}$ is the area of the corresponding\textbf{\ }outer surface of
the PNS\textbf{\ }shell of significant gradient.

To get an estimate we write down that%
\begin{equation}
\mathcal{N}^{\mathrm{gr}}\approx \frac{T_{\nu }S_{R}\left\vert g_{\mathrm{int%
}}\right\vert \left\vert a^{\prime }\right\vert }{2\left( 2\pi \right) ^{3}}.
\label{5.5}
\end{equation}%
The ratio of this value and the total numbers $\mathcal{N}_{\sigma }$ given
by Eqs.~(\ref{4.17}), (\ref{4.20}), (\ref{4.23}), and (\ref{4.26}) is 
\begin{align}
\mathcal{N}^{\mathrm{gr}}\diagup \mathcal{N}_{\sigma }\sim &e^{\pi \lambda
}T_{\nu }\left( R\Delta R\left\vert g_{\mathrm{int}}\right\vert \right)
^{-1}
\notag
\\
&\sim \left\{ 
\begin{array}{c}
10^{-7}e^{\pi \lambda }\;\mathrm{for\;compression} \\ 
10^{-6}e^{\pi \lambda }\;\mathrm{for\;neutronization}%
\end{array}%
\right. ,  \label{5.6}
\end{align}%
where we use that $R\Delta R=R_{\mathrm{c}}\Delta R_{\mathrm{c}}=10~\mathrm{%
km}^{2}$ for the compression and $R\Delta R=R_{\mathrm{n}}\Delta R_{\mathrm{n%
}}=10^{3}~\mathrm{km}^{2}$ for the neutronization. Thus, despite the fact
that the vacuum instability effects caused by the PNS shells of density
gradient are very pronounced for the neutrinos with the subcritical masses
(in this case $\lambda \lesssim 1$), they are negligible during the initial
stages of the PNS evolution and cannot block the $\nu \bar{\nu}$ pair
creation due to the time-dependent effective potential. We note, however,
that the ratio in Eq.~(\ref{5.6}) is very sensitive to the neutrino mass. If
the mass of the lightest neutrino is sufficiently greater than the critical
values given by Eqs.~(\ref{mcrbounce}), (\ref{5.1}), and (\ref{5.2}), $%
\lambda \gg 1$, so that the ratio (\ref{5.6}) is not small, $\mathcal{N}%
^{crust}\diagup \mathcal{N}_{\sigma }\gtrsim 1$, then the effects caused by
the density gradients must be taken into account. Thus, our mechanism of the 
$\nu \bar{\nu}$\ pair creation is valid if $\lambda \lesssim 1$.

The nonzero difference between the effective density $g_{\mathrm{int}}$ in
the region of creation and $g_{\mathrm{ext}}\approx 0$ in the space outside
this region affects the results of the $\nu \bar{\nu}$ pair creation due to
the time-dependent effective potential for a distant observer. To see that
we consider the radial motion of neutrinos and antineutrinos through the PNS
shells of density gradient, assuming that $\mathbf{p}_{\bot }\approx 0$.
Using the Dirac equation (\ref{eq:Direq}) with the matter density $g\left(
r\right) $, we see that, in general, the helicity is not conserved when a
neutrino moves in the inhomogeneous matter. However, if $\mathbf{p}_{\bot
}\approx 0$ the projection of the spin on the radial direction is conserved.
Note that this projection is not related to the direction of the momentum
vector then the helicity is not necessary conserved anyway. The total energy
of particles and antiparticles $p_{0}^{\left( \pm \right) }$ is conserved.
Using Eq.~(\ref{eq:Direq}), one can elaborate the following asymptotic
dispersion relations for a given value of $p_{0}^{\left( \pm \right) }$:%
\begin{align}
p_{0}^{\left( \pm \right) } =&\frac{g_{\mathrm{int}}}{2}\pm \mathcal{E}_{%
\mathrm{int}},
\quad
\mathcal{E}_{\mathrm{int}}=\sqrt{m^{2}+p_{\mathrm{int}%
}^{2}}
\notag
\\
&\mathrm{in\quad the\quad region\quad of\quad creation};  \notag \\
p_{0}^{\left( \pm \right) } =&\pm \mathcal{E}_{\mathrm{ext}},
\quad
\mathcal{%
E}_{\mathrm{ext}}=\sqrt{m^{2}+p_{\mathrm{ext}}^{2}}
\notag
\\
&\mathrm{outside\quad
the\quad region\quad of\quad creation}.  \label{5.7}
\end{align}%
Here $\mathcal{E}_{\mathrm{int}}$, $\mathcal{E}_{\mathrm{ext}}$ are the
corresponding asymptotic values of the particle kinetic energy and $p_{%
\mathrm{int}}$, $p_{\mathrm{ext}}$ are the magnitudes of the corresponding
radial momenta $p_{\mathrm{int}}=\left\vert \mathbf{p}_{\mathrm{int}%
}\right\vert $, $p_{\mathrm{ext}}=\left\vert \mathbf{p}_{\mathrm{ext}%
}\right\vert $, respectively.

Assuming that $g_\mathrm{int}=g\left( t\right) $\ in the region of creation
varies adiabatically from $g(t_{\mathrm{in}})$\ \ to $g(t_{\mathrm{out}})$,
we consider the case when $\mathcal{E}_\mathrm{int}$\ is the energy of
neutrino or antineutrino with a given $\sigma $\ created from vacuum by the
neutronization until the time $t$, $\mathcal{E}_\mathrm{int}=\mathcal{E}_{2}$%
, where $\mathcal{E}_{2}$\ is given by Eq.~(\ref{3.10}) at $g_{2}=g_\mathrm{%
int}$. Then $p_\mathrm{int}=\left\vert p-\sigma g_\mathrm{int}/2\right\vert $%
\ and the ranges of momentum are given by Eqs.~(\ref{4.19}) and (\ref{4.22})
at $g_{2}=g_\mathrm{int}$. Taking into account the fact that the main
fraction of the $\nu \bar{\nu}$-pairs with $\sigma =-1$\ for all flavors are
created due to the neutronization at the time $t$\ when%
\begin{align*}
g_\mathrm{int}<&0;\mathrm{\quad }\left\vert g_\mathrm{int}\right\vert >\sqrt{%
2a}K\mathrm{\quad for\;}\nu _{e}\bar{\nu}_{e},
\\
\left\vert g_%
\mathrm{int}\right\vert >&\left\vert g(t_{\mathrm{in}})\right\vert +\sqrt{2a}K%
\mathrm{\quad for\;}\nu _{\mu ,\tau }\bar{\nu}_{\mu ,\tau },
\end{align*}%
\textrm{\ }we find from Eq.~(\ref{5.7}) that all of these neutrinos are
bounded in the PNS while all of these antineutrinos gain additional kinetic
energy $\sim \left\vert g_\mathrm{int}\right\vert /2$\ and escape the PNS
with the energy $\mathcal{E}_\mathrm{ext}\mathcal{\approx }\left\vert g_%
\mathrm{int}\right\vert -p$. It is consistent with the general conclusion
obtained earlier for neutron stars in Refs.~\cite{Loeb90,Kach98,KiersW97}.
For all $\nu _{\mu ,\tau }\bar{\nu}_{\mu ,\tau }$\ pairs created due the
compression we have $\sigma =-1$ and $g_\mathrm{int}<0$.\ \textrm{\ }Note
that the projection of the kinetic momentum on the direction of the momentum
of this antineutrino, $p-\left\vert g_\mathrm{int}\right\vert <0$, then its
physical helicity outside the region of creation is negative. Such kind of
antineutrino does not substantially interact with the matter of electrons
and baryons, unless it interacts with a potential barrier, then it is
considered undetectable. The final effective density $g_\mathrm{int}=g(t_{%
\mathrm{out}})$\ retains its value during the entire period of the existence
of a neutron star then these neutrinos are the trapped forever. Thus, we
estimate the time-depending range of the antineutrino kinetic energy outside
the PNS during the neutronization as follows\textrm{\ }%
\begin{align}
&\frac{\left\vert g_\mathrm{int}\right\vert }{2}<\mathcal{E}<\left\vert g_%
\mathrm{int}\right\vert \mathrm{\quad for\;}\bar{\nu}_{e},
\notag
\\%
&\frac{1}{2}\left( \left\vert g(t_{\mathrm{in}})\right\vert +\left\vert g_%
\mathrm{int}\right\vert \right) <\mathcal{E}<\left\vert g_\mathrm{int}%
\right\vert \mathrm{\quad for\;}\bar{\nu}_{\mu ,\tau }.  \label{5.8}
\end{align}%
The range of the $\bar{\nu}_{\mu ,\tau }$\ kinetic energy outside the PNS
during the compression\ is%
\begin{equation}
\frac{\left\vert g_\mathrm{int}\right\vert }{2}<\mathcal{E}<\left\vert g_%
\mathrm{int}\right\vert .  \label{5.8b}
\end{equation}%
When the neutronization stage ended, the spherical layer of
ultrarelativistic antineutrinos with the kinetic energies in the range%
\begin{align}
&\sqrt{a/2}K<\mathcal{E}<\left\vert g(t_{\mathrm{out}})\right\vert \mathrm{%
\quad for\;}\bar{\nu}_{e},
\notag
\\
&\left\vert g(t_{\mathrm{in}%
})\right\vert +\sqrt{a/2}K<\mathcal{E}<\left\vert g(t_{\mathrm{out}%
})\right\vert \mathrm{\quad for\;}\bar{\nu}_{\mu ,\tau }  \label{5.9}
\end{align}%
is formed outside the PNS and then expands at\emph{\ }a speed close to the
speed of light.\textrm{\ }When the compression stage ended, the spherical
layer of $\bar{\nu}_{\mu ,\tau }$\ with the kinetic energies in the range%
\begin{equation}
\sqrt{a/2}K<\mathcal{E}<\left\vert g(t_{\mathrm{out}})\right\vert
\label{5.9b}
\end{equation}%
is formed outside the PNS and then expands.

We point out first that for the part of the $\nu _{e}\bar{\nu}_{e}$ pairs
created with the helicity quantum number $\sigma =+1$ due to the
neutronization, the effect of the PNS border is completely different. It was
shown in Eq.~(\ref{4.19}) that such particles are created before the
effective density $g\left( t\right) >0$ passes through zero at some time $%
t_{0}$ and have the maximal kinetic energy per particle $\sim \frac{1}{2}%
g(t_{\mathrm{in}})$ at $t_{0}$. Therefore, the positive value $g_\mathrm{int}%
=g(t)$ varies from $g(t_{\mathrm{in}})$\ \ to zero, meanwhile the maximal
kinetic energy of created $\nu _{e}$ or $\bar{\nu}_{e}$ increases from zero
to $\frac{1}{2}g(t_{\mathrm{in}})$. If $p>g_\mathrm{int}/2$, then both $\nu
_{e}$ and $\bar{\nu}_{e}$ escape the PNS and the time-depending range of the
kinetic energy outside the PNS during the neutronization is\emph{\ }%
\begin{align}
&\sqrt{a/2}K<\mathcal{E}<g(t_{\mathrm{in}})/2\mathrm{\quad for\;}\nu _{e},%
\notag
\\
&0<\mathcal{E}<g(t_{\mathrm{in}})/2-g_\mathrm{int}\mathrm{%
\quad for\;}\bar{\nu}_{e}.  \label{5.10}
\end{align}%
\emph{\ }Their helicity quantum number outside the crust is conserved. Such
a fraction of the $\nu _{e}\bar{\nu}_{e}$\ is considered undetectable
directly.\textbf{\ }

If $p<g_\mathrm{int}/2$\ and $g_\mathrm{int}>\sqrt{a/2}K$, then these $\bar{%
\nu}_{e}$ are bounded in the PNS until the time when $g_\mathrm{int}$\ will
be small enough and then escape\emph{\ }with helicity conserved.\emph{\ }All
of these $\nu _{e}$ gain additional kinetic energy $\sim g_\mathrm{int}/2$\
and escape the PNS with the energy $\mathcal{E}_\mathrm{ext}\approx g_%
\mathrm{int}-p$. The projection of the kinetic momentum on the direction of
the momentum of this $\nu _{e}$, $p-g_\mathrm{int}<0$, then its physical
helicity outside the PNS is negative. Such neutrinos interact with the
matter of electrons and baryons and are detectable in principle. We estimate
the time-depending range of the neutrino kinetic energy outside the region
of creation during the neutronization as%
\begin{equation}
g_\mathrm{int}/2<\mathcal{E}<g_\mathrm{int}-\sqrt{a/2}K.  \label{5.11}
\end{equation}%
This range shrinks to the point when time $t$\ tend to $t_{0}$. As a result,
when the neutronization stage ended, the spherical layer of such
ultrarelativistic neutrinos with the kinetic energies in the range%
\begin{equation}
0<\mathcal{E}<g(t_{\mathrm{in}})-\sqrt{a/2}K  \label{5.12}
\end{equation}%
is formed outside the PNS and then expands at\emph{\ }a speed close to the
speed of light.\textrm{\ }

Thus, only electron neutrinos of all $\nu \bar{\nu}$ pairs created during
the neutronization stage can be in principle detected directly by a distant
observer. However, note that the effective potential of a neutron star is
repulsive for the low-energy antineutrinos escaped the PNS. Then these
antineutrinos can change their helicity if reflected of a neutron star. In
general, the effective potential of a neutron star can considerably refracts
such low-energy $\nu $ and $\bar{\nu}$.

From the beginning we have neglected the influence of gravity and rotation.
However, PNS can have rather strong gravitational field and{\Large \ }rotate
rapidly. In principle these effects can influence the creation of $\nu \bar{%
\nu}$ pairs and their subsequent evolution especially since energies of
particles are small. For example, as was found in Ref.~\cite{Stu08}, very
low-energy antineutrinos can be captured inside a rotating PNS. The
characteristic length scale associated with gravity or rotation of PNS is in
the km range. Indeed, it can be a gravitational radius which is several km
for a PNS with the mass in the solar range. The energy corresponding to such
a length scale is $\sim (10^{-10}-10^{-9})\,\text{eV}$. In our situations
the typical energies of $\nu \bar{\nu}$ pairs are up to several $\mathrm{eV}$
or up to $0.1~\text{eV.}$ Thus gravity and rotation can affect only very
narrow part near the bottom of the spectrum of $\nu \bar{\nu}$ pairs
created. Nevertheless gravity can influence the propagation of created
neutrino beam while it propagates further in space. By the same reason a
cosmic neutrino background, expected at $1.95~\mathrm{K\sim }0.17~\mathrm{meV%
}$, is irrelevant for the case under consideration.

The coherent $\nu \bar{\nu}$ pairs creation discussed in our work is not
influenced by the pairs creation by nucleon-nucleon bremsstrahlung. Indeed,
using the results of Ref.~\cite{Raf96} one gets that $\nu \bar{\nu}$ pairs
created in nucleon-nucleon bremsstrahlung have energy $\sim 1\,\text{MeV}$
in nuclear matter with temperature $T\sim 10^{9}\,\text{K}$, which is
typical for a core collapsing supernova. Thus this process does not overlap
with the pairs creation by our mechanism.

\section{SOME\ PROPERTIES\ OF\ WEBER PARABOLIC CYLINDER FUNCTIONS\label{App}}

In this appendix we list some properties of the WPCFs which are used in the
present work and where already used by us studying particle creation from
the vacuum by a quasiconstant uniform electric field, see Ref.~\cite{GavG96}.

The solution of the ordinary differential equation 
\begin{equation}
\left[ \frac{\mathrm{d}^{2}}{\mathrm{d}z^{2}}+\rho +\frac{1}{2}-\frac{z^{2}}{%
4}\right] \varphi \left( z\right) =0,  \label{eq:fzeq}
\end{equation}%
can be expressed as a linear combination of any of two functions from the
following set: $D_{\rho }(z)$, $D_{\rho }(-z)$, $D_{-\rho -1}(\mathrm{i}z)$,
and $D_{-\rho -1}(-\mathrm{i}z)$. If we change the variable $z=(1-i)\xi $ in
Eq.~(\ref{eq:feq}), we can represent it in the form of Eq.~(\ref{eq:fzeq})
with $\rho =\mathrm{i}\lambda /2+\left[ \chi \mathrm{sgn}(a)-1\right] /2$.
Then assuming that $\chi \mathrm{sgn}(a)=+1$, we obtain linear independent
solutions of Eq.~(\ref{eq:feq}) used in Sec.~\ref{S4}. Note that a more
detailed description of the properties of the WPCFs can be found, e.g., in
Ref. \cite{BatE53}.

The asymptotic expansions of WPCF, used in Sec.~\ref{S4}, corresponding to
the great absolute values of the argument $\left\vert \xi \right\vert $,
have the following form:%
\begin{align}
D_{\rho }[(1\pm \mathrm{i})\xi ]=&e^{\mp \mathrm{i}\xi ^{2}/2}\left( \sqrt{2}%
e^{\pm \mathrm{i}\pi /4}\xi \right) ^{\rho }
\notag
\\
&\times
\left[ 1\mp \mathrm{i}\frac{\rho
\left( 1-\rho \right) }{4\xi ^{2}}+\ldots \right]
\notag
\\
&\mathrm{if}\quad \xi
\geq K,  \label{a1}
\end{align}%
where $K\gg \max \left\{ 1,\lambda \right\} $. If $\xi <0$ one gets that%
\begin{align}
D_{\rho }[(1-\mathrm{i})\xi ]=&e^{\mathrm{i}\pi \rho }D_{\rho }[(1-\mathrm{i%
})\left\vert \xi \right\vert ]
+\mathrm{i}\frac{\sqrt{2\pi }}{\Gamma (-\rho )}%
e^{\mathrm{i}\pi \rho /2}
\notag
\\
&\times
D_{-\rho -1}[(1+\mathrm{i})\left\vert \xi
\right\vert ],  \notag \\
D_{-\rho -1}[(1+\mathrm{i})\xi ]=&e^{\mathrm{i}\pi \left( \rho +1\right)
}D_{-\rho -1}[(1+\mathrm{i})\left\vert \xi \right\vert ]
\notag
\\
&-\mathrm{i}\frac{%
\sqrt{2\pi }}{\Gamma (\rho +1)}e^{\mathrm{i}\pi \left( \rho +1\right)
/2}
\notag
\\
&\times
D_{\rho }[(1-\mathrm{i})\left\vert \xi \right\vert ],  \label{a2}
\end{align}%
where $\Gamma (z)$ is the Euler gamma function.

Using Eqs.~(\ref{a1}) and (\ref{a2}), we get the expansions of the
coefficients $f_{k}(t_{l})$, $k,l=1,2$, which are required for the
calculation of the expression $B$ given by Eq.~(\ref{4.9}), 
\begin{align}
f_{1}(t) \approx &O\left( \xi ^{-3}\right) ,  \notag \\
f_{2}(t) \approx &e^{\mathrm{i}\xi ^{2}/2}\left( \sqrt{2}e^{-\mathrm{i}\pi
/4}\xi \right) ^{\rho }\left[ 2+O\left( \xi ^{-2}\right) \right]
\notag
\\ 
&\mathrm{if}\quad \xi \geq K;  \notag \\
f_{1}(t) \approx &e^{\mathrm{i}\pi \left( \rho +1\right) }e^{-\mathrm{i}\xi
^{2}/2}\left( \sqrt{2}e^{\mathrm{i}\pi /4}\left\vert \xi \right\vert \right)
^{-\rho -1}
\notag
\\
&\times\left[ 2+O\left( \left\vert \xi \right\vert ^{-2}\right) \right] ,
\notag \\
f_{2}(t) \approx &O\left( \left\vert \xi \right\vert ^{-1}\right) \quad 
\mathrm{if}\quad \xi <0,\;\left\vert \xi \right\vert \geq K.  \label{a4}
\end{align}

\end{document}